\documentclass{egpubl}
\usepackage{eg2024}
 
\ConferencePaper        
\CGFccby

\usepackage{amsmath}
\usepackage[T1]{fontenc}
\usepackage{dfadobe}  
\usepackage{multirow}
\usepackage[table,xcdraw]{xcolor}
\usepackage{float}
\usepackage{todonotes}
\usepackage{listings}
\usepackage{soul}
\usepackage{makecell}
\usepackage{amssymb}
\usepackage{tabularx}

\biberVersion

\BibtexOrBiblatex
\usepackage[backend=biber,bibstyle=EG,citestyle=alphabetic,backref=true]{biblatex} 
\electronicVersion
\PrintedOrElectronic
\PassOptionsToPackage{backref=false}{hyperref}
\usepackage{egweblnk} 

\usepackage{subcaption}
\definecolor{codegreen}{rgb}{0,0.6,0}
\definecolor{codegray}{rgb}{0.5,0.5,0.5}
\definecolor{codepurple}{rgb}{0.58,0,0.82}
\definecolor{backcolour}{rgb}{0.95,0.95,0.92}

\lstdefinestyle{mystyle}{
    backgroundcolor=\color{backcolour},   
    commentstyle=\color{codegreen},
    keywordstyle=\color{magenta},
    numberstyle=\tiny\color{codegray},
    stringstyle=\color{codepurple},
    basicstyle=\ttfamily\small,
    breakatwhitespace=false,         
    breaklines=true,                 
    captionpos=b,                    
    keepspaces=true,                 
    numbers=left,                    
    numbersep=5pt,                  
    showspaces=false,                
    showstringspaces=false,
    showtabs=false,                  
    tabsize=2
}

\usetikzlibrary{shadows.blur}

\title[Variable-Rate Texture Compression: Real-Time Rendering with JPEG]%
{Variable-Rate Texture Compression: \\ Real-Time Rendering with JPEG}

\author[M. Kristmann \& M. Wimmer \& M. Schütz]
    {\parbox{\textwidth}{\centering Elias Kristmann, Michael Wimmer, Markus Schütz
    }
    \\
    {\parbox{\textwidth}{\centering TU Wien}}
}

\begin{document}

\teaser{
 \includegraphics[width=0.9\linewidth]{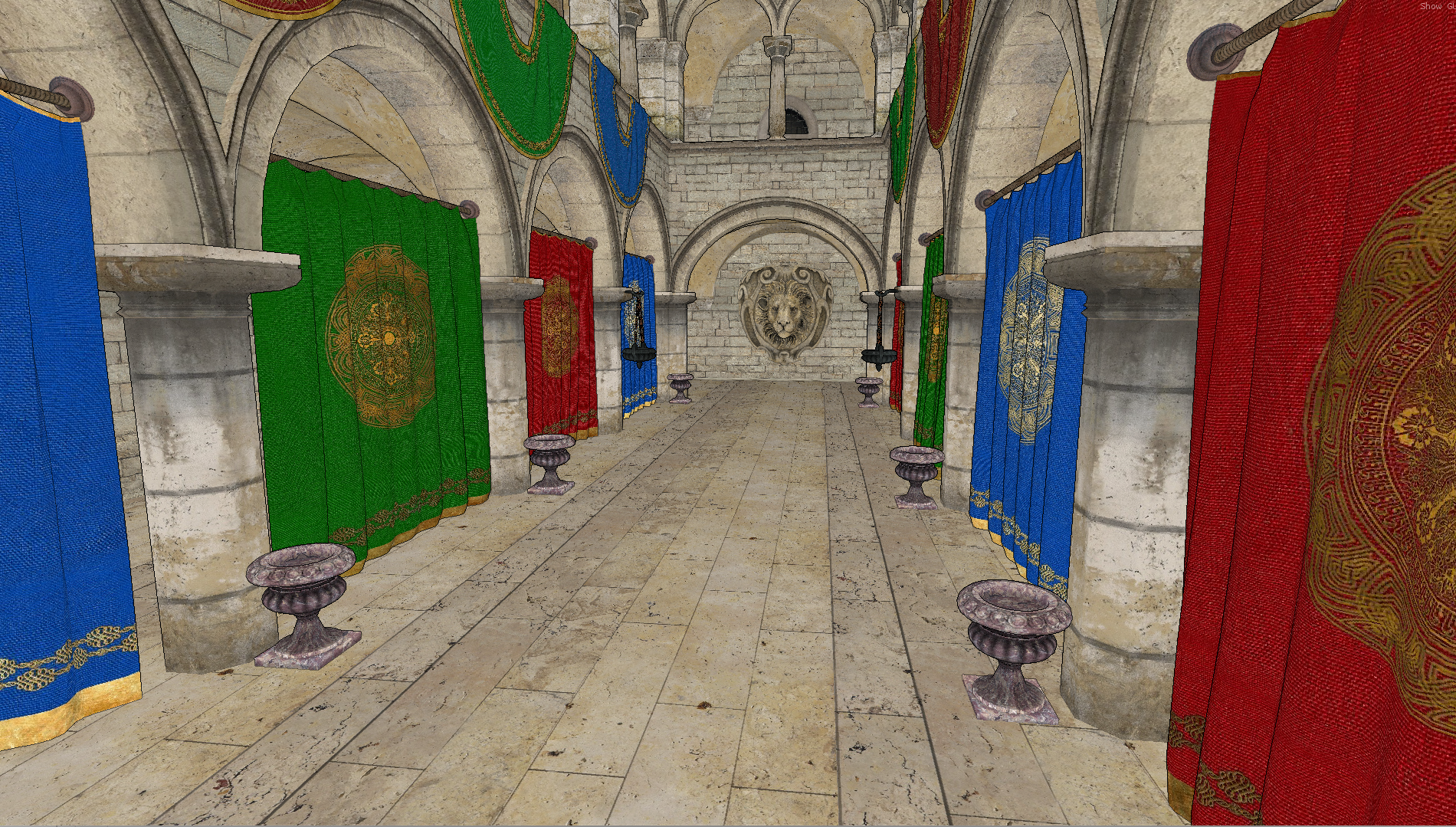}
 \centering
  \caption{Rendering Sponza with JPEG-compressed textures at 1500+ frames per second.}
\label{fig:teaser}
}

\maketitle
\begin{abstract}
Although variable-rate compressed image formats such as JPEG are widely used to efficiently encode images, they have not found their way into real-time rendering due to special requirements such as random access to individual texels. 
In this paper, we investigate the feasibility of variable-rate texture compression on modern GPUs using the JPEG format, and how it compares to the GPU-friendly fixed-rate compression approaches BC1 and ASTC. Using a deferred rendering pipeline, we are able to identify the subset of blocks that are needed for a given frame, decode these, and colorize the framebuffer's pixels. Despite the additional $\sim$0.17 bit per pixel that we require for our approach, JPEG maintains significantly better quality and compression rates compared to BC1, and depending on the type of image, outperforms or competes with ASTC. The JPEG rendering pipeline increases rendering duration by less than 0.3 ms on an RTX 4090, demonstrating that sophisticated variable-rate compression schemes are feasible on modern GPUs, even in VR.

Source code and data sets are available at: \url{https://github.com/elias1518693/jpeg\_textures}


\end{abstract}  

\section{Introduction}
In the rapidly evolving landscape of digital media, few technologies have endured with as much prominence as JPEG \cite{125072}. However, although it remains a popular image \emph{storage} format, JPEG -- and other variable-rate compressed image formats -- are considered unsuitable for real-time \emph{rendering} since they do not allow efficient access to individual texels. Due to the variable bit rates of different sections of the image, we cannot easily map a texel's index or uv-coordinates to its location in memory without creating additional indexing tables that raise the texture's memory usage. Furthermore, JPEG is based on Huffman-coded~\cite{4051119} data that must be decoded sequentially, and therefore limits our ability to parallelize the decompression. Finally, pixels in JPEG are encoded block-wise in the frequency domain via coefficients of discrete cosine transformations, requiring numerous multiplications and additions to retrieve a single pixel's color. 

In contrast, GPU-friendly fixed-rate compression algorithms encode texels in blocks with equal bit rates, allowing us to directly map a texel's coordinate to its location memory. We can then use the per-block encoded data such as color gradients and per-texel encoded data such as the weight to efficiently decode the color value with little computational effort. However, this comes at a cost of reduced compression efficiency. With limited VRAM capacity and the substantial increase in texture data associated with modern photorealistic rendering, these fixed-rate compression techniques are approaching the point where they can no longer meet the memory requirements.

JPEG and modern counterparts such as AVIF, WebP, and JPEG XL achieve higher compression ratios, enabling larger amounts of high-resolution textures to be stored in GPU memory while also reducing disk-to-memory streaming times. In this paper, we investigate whether the principles of variable-rate image compression can be applied effectively in real-time rendering. We focus on JPEG, arguably the most widely used variable-rate format. By demonstrating its feasibility for this purpose, we aim to challenge prevailing assumptions about the unsuitability of variable-rate compression for GPU textures and to lay the groundwork for adopting more advanced variable-rate compression schemes in real-time rendering.

In particular, our contributions to the state-of-the-art are:

\begin{itemize}
    \item An efficient deferred rendering pipeline capable of real-time rendering JPEG-compressed textures with an overhead of less than 0.3ms on an RTX 4090.
    \item A texture block cache that retains previously decoded JPEG blocks, reducing the amount of data that needs to be decoded in each frame.
    \item A performance study that demonstrates massive benefits of mip mapping for this approach due to reduced decoding workload; the efficiency of caching; and the suitability to VR-rendering since stereo viewpoints share most of the required texels.
    \item A quality study that illustrates the benefits of variable-rate over fixed-rate texture compression, particularly AVIF and JPEG XL. 
\end{itemize}

\section{Related Work}

In the context of real-time-rendering, we can categorize texture compression algorithms into fixed-rate compression, variable-rate compression, and the emerging neural texture compression approaches. 


\subsection{Fixed-rate compression}

Fixed-rate compression refers to the fact that all segments of the image are encoded with the same bit rate, which allows us to compute the memory location of a texel from its coordinate. GPU-based algorithms also make an effort to keep the decoding computationally inexpensive. 

Despite being over two decades old, S3 Texture Compression (S3TC) \cite{Iourcha1999S3TC}, also known as DXT or more recently BC1 through BC7, remains one of the most widely used texture compression algorithms today. BC1, for example, compresses images by dividing them into 4×4 pixel blocks and computing two color endpoints in RGB space for each block. The algorithm then encodes all pixels within the block as interpolations between these endpoints. BC1 and BC7 are tailored for color images, with BC1 offering lower quality but achieving twice the compression rate at 4 Bits per Texel (bpt) compared to the more modern and higher-fidelity BC7 at 8 bpt. Similarly, Adaptive scalable texture compression (ASTC) \cite{10.2312:hpg.20211284} builds on this idea but makes the block sizes adaptive. This allows the method to adjust better to the characteristics of a texture, achieving higher and more controllable compression, from 0.89 to 8.0 bpt, with even better visual quality. However, this improved compression comes with an increased complexity. Ericsson Texture Compression (ETC1/ETC2), available in OpenGL ES, targets with its low complexity lower-end devices and mobile phones \cite{10.1145/1071866.1071877}. More recently, Neural Texture Block Compression \cite{https://doi.org/10.2312/mam.20241178} uses a neural network that learns to optimize BC1 compression and therefore achieves better results for the same storage format than the handcrafted compression algorithms. Chen et al. propose a form of fixed-rate JPEG where they compress each JPEG block with different quantization tables until it fits into the fixed block size \cite{Chen2002AJT}. Hollemeersch et al. transform a texture with the discrete cosine transform (DCT) and then only store a fixed and limited number of coefficients for each block \cite{10.1007/s00371-011-0621-8}. While this only slightly improves the compression rate compared to BC1, they also show that texture filtering can be done in the frequency domain before decoding, greatly increasing its efficiency.

\subsection{Variable-rate compression}

Variable-rate compression formats adjust the bit rate to the content, making some sections of an image compress better than others. Since they do not cater to real-time rendering, they also employ complex and computationally expensive algorithms that sacrifice decoding performance for higher compression rates. 

The most famous lossy compression standard for images is JPEG~\cite{125072} defined in 1992. Since then, many follow-ups have emerged, offering new features and improved compression, such as JPEG2000~\cite{10.1145/357744.357757} and most recently JPEG XL~\cite{48554, sneyers2025jpegxlimagecoding}. However, despite their advancements, these newer algorithms have not yet achieved the widespread support and adoption of the original JPEG standard. 

Random access to JPEG blocks has been successfully applied to offline rendering, as shown by  Radziszewski et al. \cite{radziszewski2008optimization}. Because the algorithm is specifically designed for CPU-based processing, its decoding efficiency is relatively limited. However, it provides a straightforward method for accessing individual texels in JPEG-compressed images, by maintaining a list of offsets to the start of each Minimum Coded Unit (MCU) in the compressed data. Olano et al. introduce an online variable-rate texture compression technique that also stores an index list to each MCU and uses it to parallelize and therefore speed up the decoding process \cite{10.1111:j.1467-8659.2011.01989.x}. They avoid problems related to the missing random access by completely decompressing the textures on the GPU before rendering. Consequently, this approach provides storage savings primarily in scenarios where only a subset of textures is required at any given time, allowing for more textures to be stored in memory, but it does not reduce VRAM usage for an individual texture while it is used. 
Although far from a real-time application, Fichet et al. demonstrate a method for compressing spectral images used in spectral rendering by converting them to the JPEG XL format, and then decoding them while rendering, achieving file size reductions of 10 to 60 times compared to ZIP compressed files \cite{fichet2025compression}.

\subsection{Neural texture compression}
 Recently, the trend has shifted to neural texture compression methods, utilizing neural networks to learn efficient and perceptually optimized compression schemes from feature vectors. These methods often achieve higher compression ratios and improved visual quality compared to conventional block-based compression algorithms, but have reduced decoding efficiency as a tradeoff. In  Random-Access Neural Compression of Material Textures (NTC) \cite{ntc2023}, the textures are stored as a feature pyramid, with multiple channels compressed together. Therefore, this method excels in scenarios with various channels that have a strong correlation. Farhadzadeh et al. \cite{farhadzadeh2024neuralgraphicstexturecompression} promise even greater compression rates with their neural compression method, but offer no indication of the decoding efficiency of their method. 
Weinreich et al. \cite{weinreich2024realtimeneuralmaterialsusing} leverage existing hardware support for S3TC block compression to encode and store their neural texture compression network, thereby reducing overall memory requirements. Building on this approach, Belcour et al. \cite{laurent2025hardwareacceleratedneuralblock} enhance the performance by utilising cooperative vectors.

\subsection{Other}

Schuster et al. ~\cite{10.2312:hpg.20211284} introduce a sparse coding technique to efficiently represent textured splats. For each splat data set, they construct a dictionary of "atoms" -- small images (e.g. 32x32 pixels) that capture frequently occuring patterns in a data set. Each splat holds an average color, a list of indices to atoms, and a set of weights with which the atoms are multiplied to recover the texture.  Luo et al. reduce redundancy by merging all the textures in a scene and remapping the texture coordinates to eliminate repeated content \cite{10.1145/3610548.3618150}.
Zhang et al. \cite{zhang2024gaussianimage1000fpsimage} propose using 2D Gaussian splats — a representation recently popularized in photorealistic 3D rendering \cite{kerbl3Dgaussians} — for image representation and compression. To further improve compression they apply entropy encoding to the Gaussians. Zhang et al. \cite{zhang2025image} propose a similar approach but manage to improve the results even without entropy coding. Their method adaptively distributes Gaussians based on local image complexity, allocating more splats to regions with fine structural detail. This leads to higher reconstruction fidelity in edge-rich areas while maintaining efficient compression overall. The representation supports random access, making it a possible candidate for texture compression.\\

\section{Background}

This section presents an overview of the concepts and algorithms behind JPEG and BC1, providing the fundamentals to understand our JPEG texture mapping approach, as well as reasons why standard JPEG is generally considered unsuitable for GPUs, and why BC1 is deemed suitable. 

\subsection{JPEG}

JPEG is built on several concepts such as Huffman coding, difference coding, encoding in the frequency-domain, human perception, quantization, etc., each of which contributes to the compression efficiency. The YCbCr color space is used instead of RGB as it allows exploiting the higher sensitivity of humans towards differences in luminance. This color space separates luminance (Y) from chrominance components: blue-difference (Cb) and red-difference (Cr). In this paper, we specifically focus on JPEGs with 4:2:0 chroma subsampling -- a commonly used configuration that samples luminance at full resolution but color information at half resolution (one Cb and Cr value per 2x2 pixels). 

\begin{figure}[]
    \begin{subfigure}[t]{0.32\linewidth}
        \includegraphics[width=\textwidth]{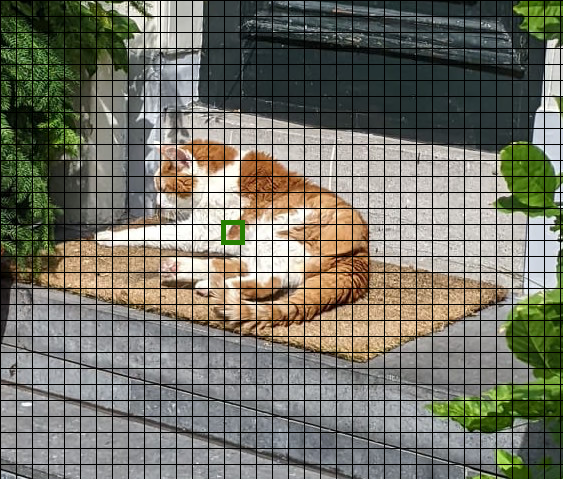}
        \caption{Image}
    \end{subfigure}
    \hfill
    \begin{subfigure}[t]{0.32\linewidth}
        \includegraphics[width=\textwidth]{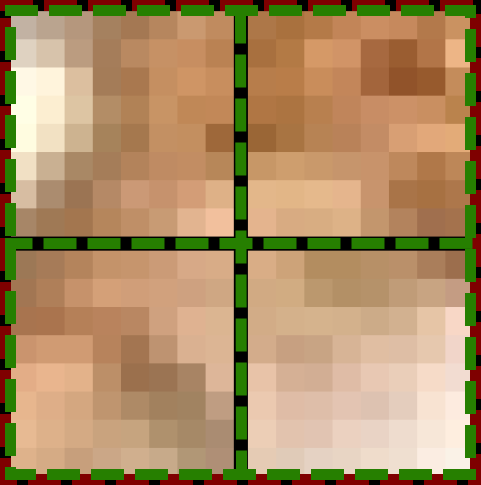}
        \caption{MCU}
    \end{subfigure}
    \hfill
    \begin{subfigure}[t]{0.32\linewidth}
        \includegraphics[width=\textwidth]{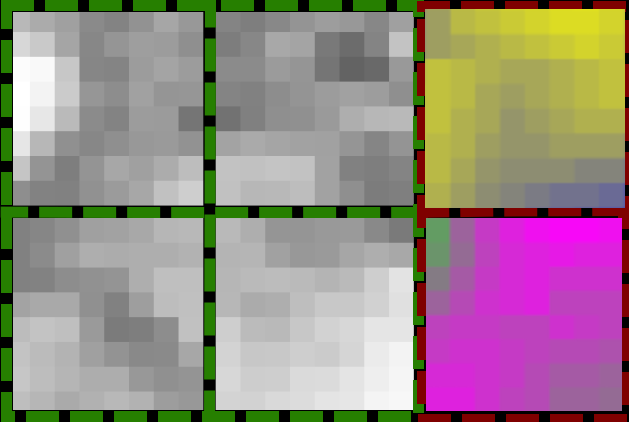}
        \caption{Data Units (DU)}
    \end{subfigure}
    \caption{JPEG images with 4:2:0 chroma sampling are split into Minimum Coded Units (MCUs; 16x16 pixels), which are encoded in six Data Units (DUs; 8x8 values). }
    \label{fig:jpeg_mcus}
\end{figure}

\textbf{Structure}: JPEG compresses images in blocks called minimum coded units (MCUs, see Figure~\ref{fig:jpeg_mcus}) -- aptly named after the fact that accessing a single pixel inside an MCU requires decoding the entire MCU. The pixel size of an MCU varies depending on the configuration. Using 4:2:0 chroma subsampling, each MCU represents a block of 16x16 pixels, and each MCU is further subdivided into 6 Data Units (DUs) that store 8x8 values. Since we sample luminance at full resolution and colors at half resolution, 4 of these 6 DUs are for luminance (one value per pixel). The remaining two are for Cb and Cr, respectively, each of which stores one value per 2x2 pixels. However, instead of storing colors per-pixel in the spatial domain, JPEG encodes colors in the frequency domain in form of a weighted sum of DCT components. The first weight/coefficient of each DU is called the DC component, from which the average pixel value is derived. The remaining 63 coefficients are the AC components. Compressing a 4096×4096 pixel image results in 65,536 MCUs and a total of 393,216 DUs.

\begin{figure*}[t]
    \centering
    \includegraphics[width=1\linewidth]{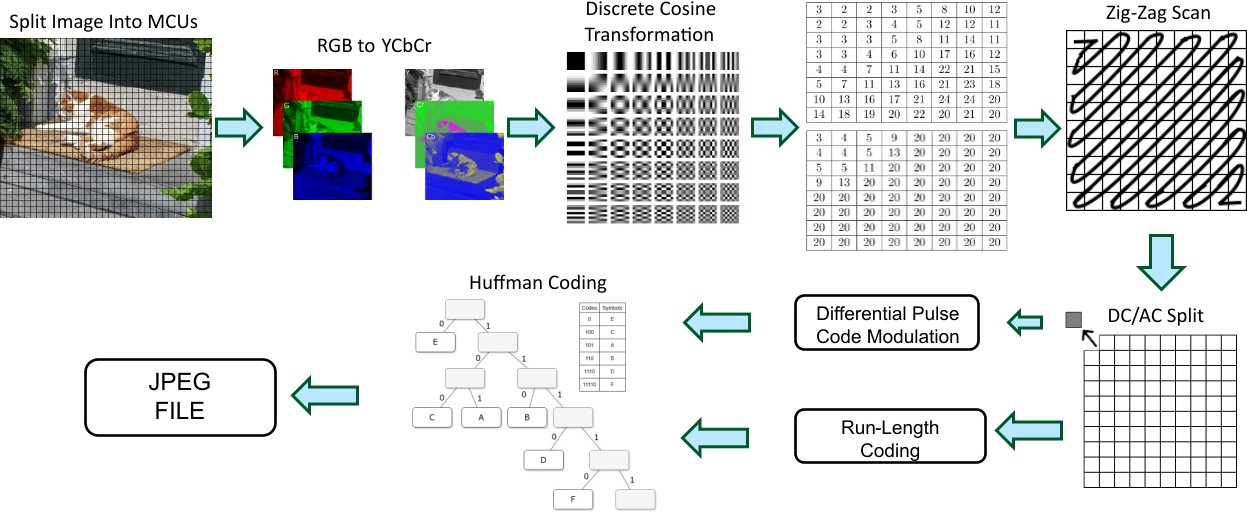}
    \caption{JPEG Compression Pipeline}
    \label{fig:jpeg_compression_pipeline}
\end{figure*}

\textbf{Pipeline}: To compress an image using baseline JPEG (Figure~\ref{fig:jpeg_compression_pipeline}), the image is first converted from the RGB color space to YCbCr. Next, chroma subsampling is applied using a 4:2:0 scheme. In this process, the Cb and Cr channels are sampled at half the horizontal and vertical resolution compared to the Y channel. The image is then divided into MCUs (16x16 pixel) and DUs (six 8x8 blocks). Each DU undergoes a Discrete Cosine Transform (DCT), representing the block as a weighted sum of 64 cosine waves with varying frequencies. The coefficient at zero frequency in both dimensions, known as the DC coefficient, corresponds to the average of all samples. To further reduce the data size, the DC coefficients are encoded differentially, meaning that each DC coefficient is stored as the difference relative to the previous one inside the same channel, rather than its absolute value. The remaining 63 non-zero frequency coefficients are called AC coefficients. The AC coefficients are then quantized, typically using a standard quantization matrix, which discards many high-frequency components, resulting in numerous zero values. The different quality settings of JPEG modify this quantization matrix, where a low quality setting discards more frequencies. Finally, the quantized coefficients are compressed using run-length encoding followed by Huffman encoding. The decoding process follows the same steps, but in reverse order. 

The major reasons why JPEG is considered unsuitable for real-time rendering on GPUs are:
\begin{enumerate}
    \item No random access to individual texels. Due to the variable bit size of each MCU, we cannot directly map uv-coordinates to the memory location of MCUs and its texels without indexing tables that raise memory usage. 
    \item Even knowing the memory location of an MCU, fetching a single pixel during rasterization is expensive, as it requires decoding the entire MCU. This is because JPEG encodes in the frequency domain, where each coefficient affects the value of all pixels in the MCU. In other words, each individual pixel is a computationally expensive dot product of 384 elements -- one dot product of 64 coefficients x 64 DCT components per DU. 
    \item AC coefficients are encoded sequentially, requiring us to decode 63 ACs per DU for a total of up to 378 ACs per MCU in a single-threaded fashion. 
    \item ACs are Huffman coded -- decoding a single AC requires a loop to find a match between the next n bits in the stream and tens of codes in the Huffman table. 
\end{enumerate}

Our approach, as described in Section~\ref{sec:method}, addresses these issues through a combination of solutions such as a compact indexing table; warp-level parallel Huffman decoding; an MCU cache that allows reusing 16x16 blocks of pixels that were decoded in previous frames; and a deferred rendering pipeline as standard forward-renderers would run into unproductive stalls while waiting until texels are decoded. 

\subsection{BC1}

The BC1 to BC7 compression formats all follow similar ideas and we will specifically focus on the BC1 format. BC1 stores textures in blocks of 4x4 pixels, using 64 bit per block for a total of 4 bit per pixel. 32 of a block's bits are used to store two 16 bit color values $c_0$ and $c_1$. The remaining 32 bit are distributed over the 16 pixels, encoding 2-bit interpolation weights between $c_0$ and $c_1$. As two bits grant us four possible states, each pixel can be either $c_0$; $\frac{2}{3}c_0 + \frac{1}{3} c_1$; $\frac{1}{3}c_0 + \frac{2}{3} c_1$; or $c_1$.

Encoding and finding suitable color endpoints that minimize the compression loss can be computationally expensive, but decoding is near-trivial. In contrast to the heavy and complex decoding pipeline of JPEG, BC1 only requires fetching 8 bytes, unpacking the two 16 bit color values, and interpolating the final color using each texel’s 2-bit weight.

\section{Method}
\label{sec:method}

\begin{figure*}[htbp]
    \centering
    \includegraphics[width=0.9\linewidth]{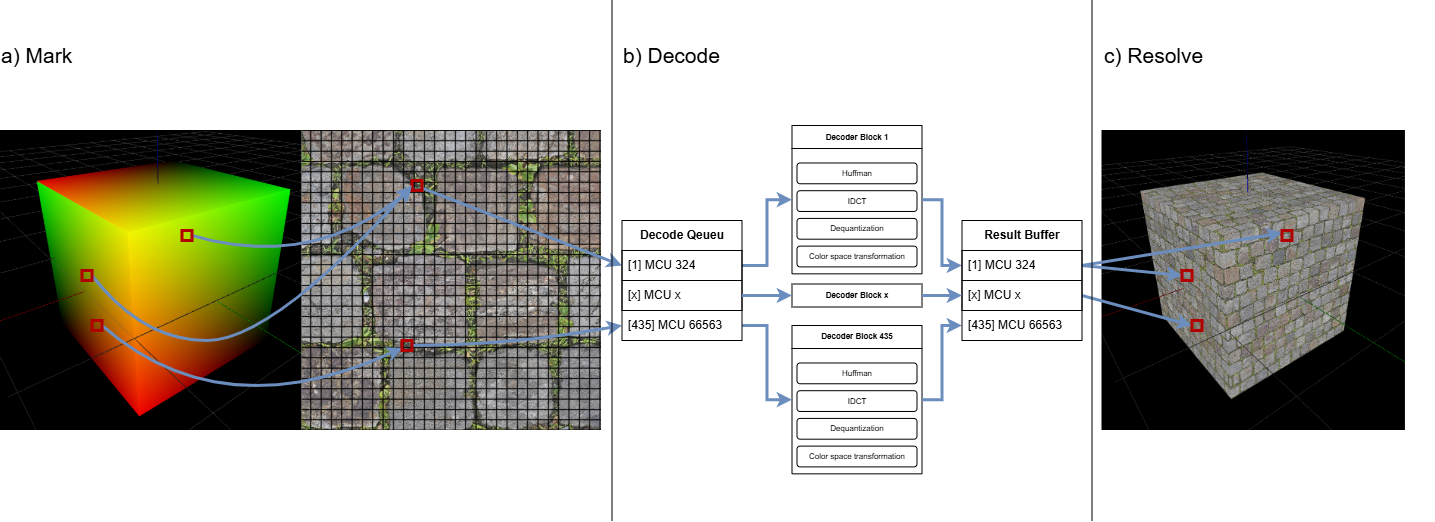}
    \caption[Overview of our method.]{ Overview of our method. a) For every pixel's UV coordinate, we compute the corresponding MCU and add it to the decoding queue. If an MCU is required multiple times, it is still added to the queue only once. b) MCUs in the decoding queue are processed in parallel by GPU blocks. Each block decodes its assigned MCU and writes the resulting data to a buffer, preserving the order of insertion in the queue. c) The decoded MCUs are retrieved from the buffer and sampled to compute the color for each pixel, completing the rendering process.}
    \label{fig:pipeline}
\end{figure*}

This section provides an algorithmic overview. Implementation details are provided in Section~\ref{sec:implementation_details}. We begin by determining which regions of each texture are actually required for rendering. Only these relevant portions are then decoded, enabling the scene to be rendered as usual without the need to decompress entire textures. Furthermore, any decoded texture regions that are needed in the subsequent frame are retained in a cache, reducing redundant decoding work and improving overall rendering efficiency. The overall process, as illustrated in Figure~\ref{fig:pipeline}, is integrated into a deferred rendering pipeline, which consists of the following passes:

\begin{enumerate}
    \item \textbf{Geometry Pass}: Draws all geometry into a G-Buffer comprising uv-coordinates, texture ID, and mip map level. 
    \item \textbf{Mark}: Screen pass that identifies the MCUs that need decoding and reserves slots in the texture block cache. 
    \item \textbf{Decode}: Decodes queued MCUs and writes texels into the texture block cache. 
    \item \textbf{Resolve}: A screen pass that maps the G-Buffer's uv-coordinates, texture IDs and mip levels to decoded texel colors in cache.
    \item \textbf{Update Cache}: Evict MCUs from cache that were not used in this frame. 
\end{enumerate}

\begin{figure}[]
    \begin{subfigure}[t]{0.49\linewidth}
        \includegraphics[width=\textwidth]{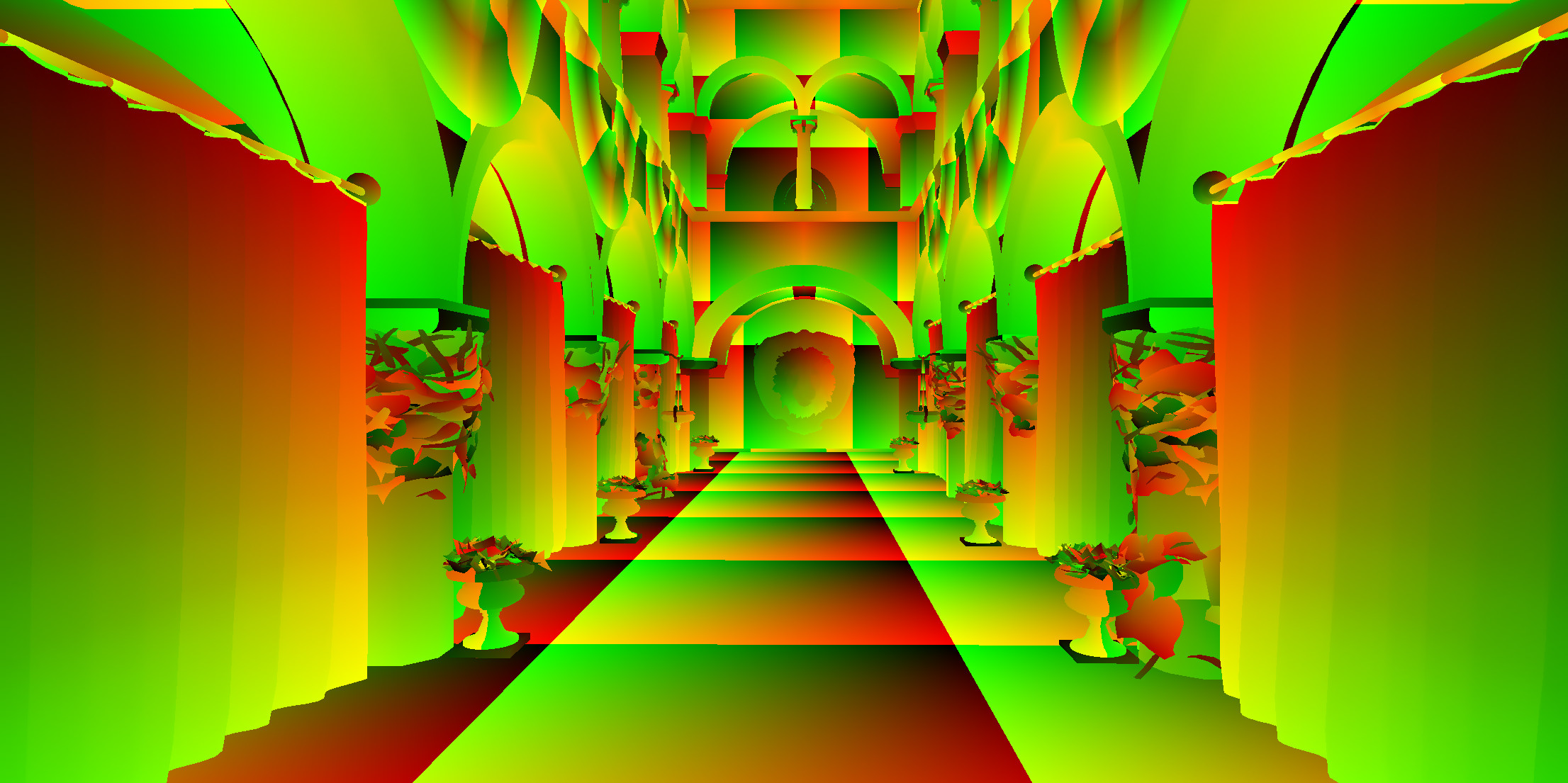}
        \caption{UVs}
    \end{subfigure}
    \hfill
    \begin{subfigure}[t]{0.49\linewidth}
        \includegraphics[width=\textwidth]{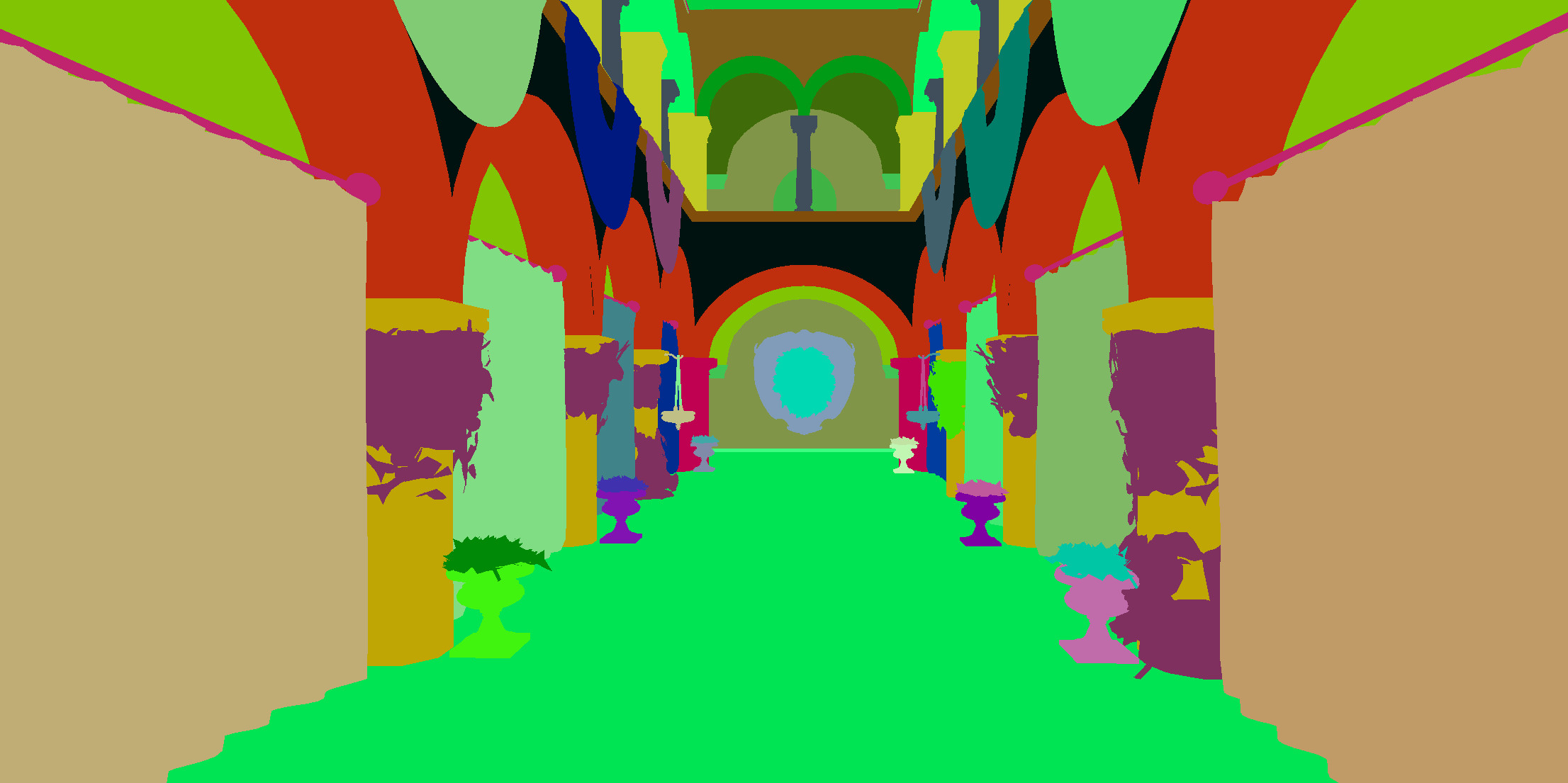}
        \caption{Texture IDs}
    \end{subfigure}
    \hfill
    \begin{subfigure}[t]{0.49\linewidth}
        \includegraphics[width=\textwidth]{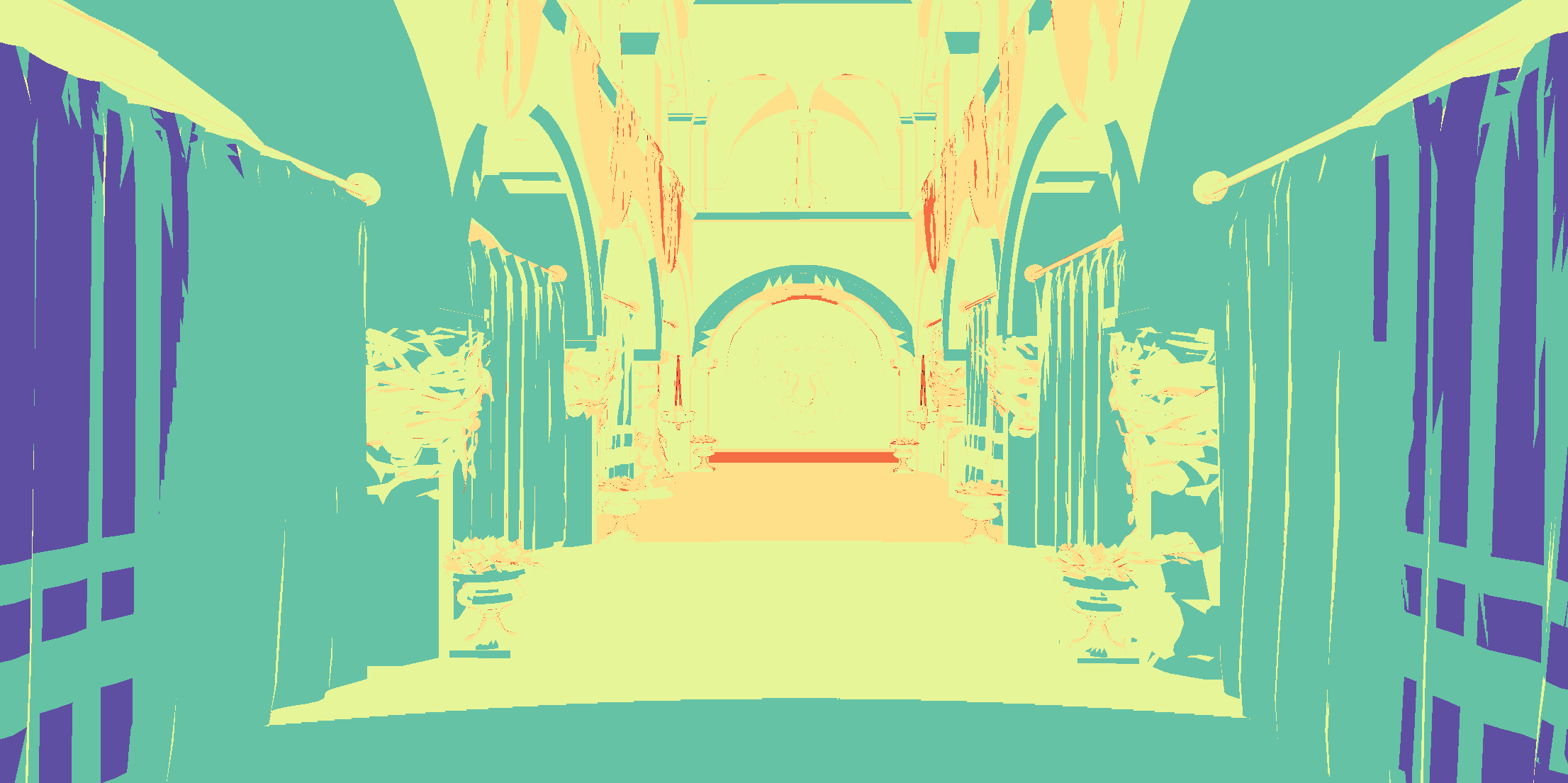}
        \caption{Mip Level}
    \end{subfigure}
    \hfill
    \begin{subfigure}[t]{0.49\linewidth}
        \includegraphics[width=\textwidth]{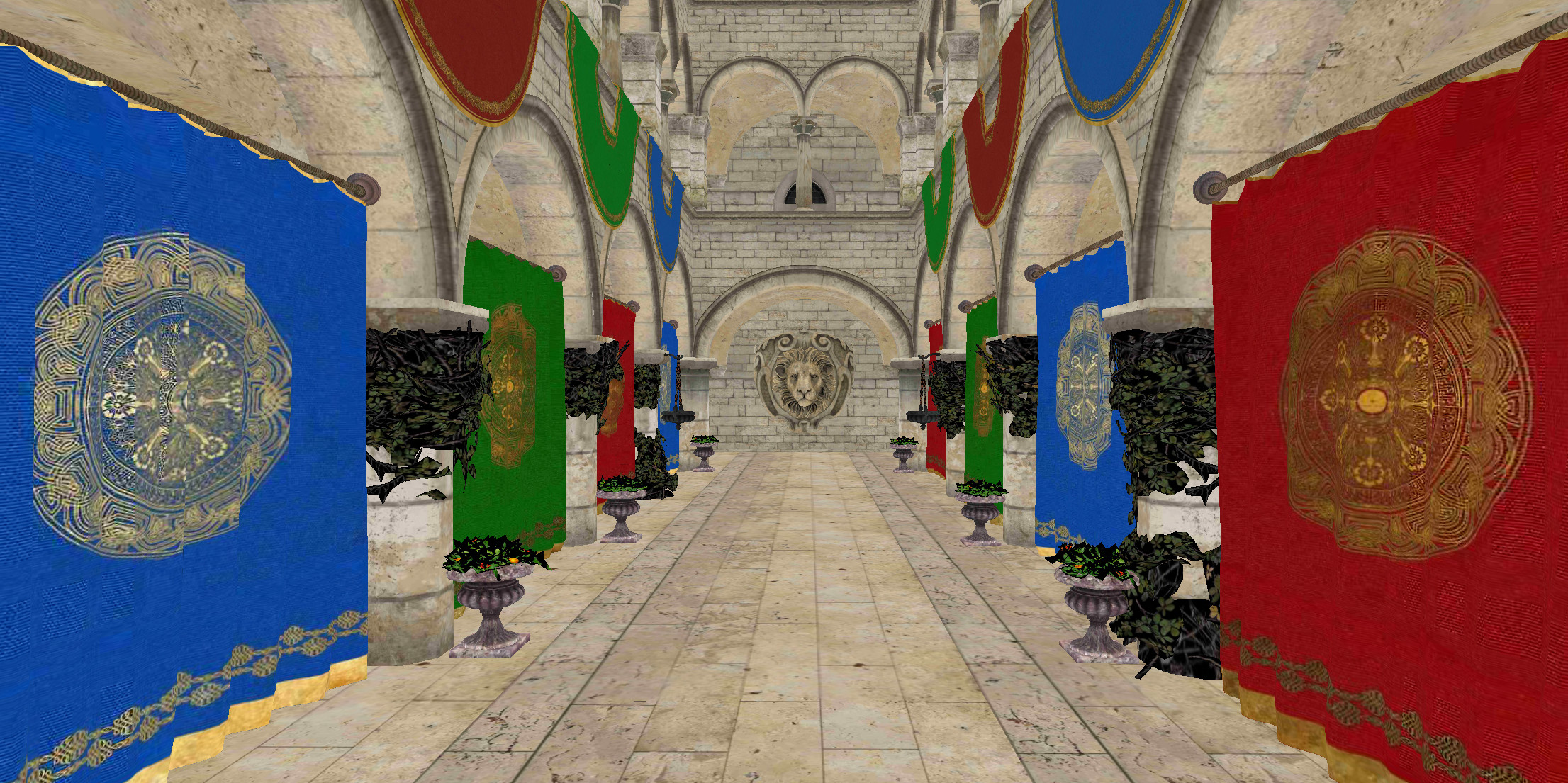}
        \caption{Rendering}
    \end{subfigure}
    \caption{(a,b,c) G-Buffer components. (d) Textured Rendering.}
    \label{fig:buffers}
\end{figure}

Figure~\ref{fig:buffers} shows the G-Buffer components from the geometry pass, and the final textured rendering after the resolve. 

\subsection{Texture Block Cache}

To avoid re-decoding all visible MCUs each frame, we implement a texture block cache. It consists of a hash map that maps the index of an MCU to the block of 16x16 decoded texels, and a pool of such blocks from which entries can be acquired or released as needed. 

\subsection{Mark}

This pass analyzes each pixel of the G-Buffer to identify the MCUs that we need to decode, but also to flag all previously decoded and still visible MCUs to keep them in cache. From the UV coordinate, we calculate the ID of the MCU:

$\mathrm{MCU} = 
\frac{(\lfloor u \cdot \mathrm{width} \rfloor \bmod \mathrm{width})}{16} 
+ \frac{(\lfloor v \cdot \mathrm{height} \rfloor \bmod \mathrm{height})}{16} 
\cdot  \lfloor \frac{\mathrm{width}}{16} \rfloor$

We then attempt to atomically reserve a spot in the cache's hash map using atomicCAS. If the MCU was not yet cached, the thread that was able to reserve a new spot in the hash map continues to add the MCU to a queue. If the MCU was already present in the cache, the thread instead sets that MCUs visibility flag inside the cache to true, ensuring that it will not get evicted from cache at the end of the frame. 

\subsection{Decode}

Next, we launch a kernel with one thread block of size 64 for each queued MCU. Since the byte offset of each MCU is stored in the indexing table, all MCUs can be decoded independently and in parallel. Each block retrieves an MCU from the decoding queue, reads its byte offset in the compressed stream from the indexing table, and loads the subsequent 384 bytes into local memory to optimize access during decoding. The next step, and the main performance bottleneck during MCU decoding, is the Huffman decoding of AC coefficients, which must be executed sequentially. However, subsequent steps, including dequantization, inverse discrete cosine transform (IDCT), and color space conversion, are highly parallelizable and well-suited for efficient GPU computation. The decoded RGB values are stored in the texture block cache by first acquiring a free 16x16 block from the cache pool, writing the texels into it, and updating the corresponding hash map entry with a handle to the block. 

\subsection{Resolve}

This pass launches one thread per pixel, converting the G-Buffer's uv-coordinate and texture index to the decoded texels. Using the MCU ID, we query the location of the decoded 16x16 block of texels from the cache via a hash map lookup, and retrieve the texel that corresponds to the uv index. 

Linear interpolation is supported by fetching the four adjacent texels and weighting them according to the G-Buffer's uv-coordinate. A potential issue may be that adjacent texels could belong to different MCUs, which may or may not be decoded. In practice we found that in the vast majority of cases, adjacent MCUs are typically also visible and decoded. If not, we clamp and duplicate the bordering texel value with little to no observable artifacts. 

\subsection{Update Cache}

After the frame is rendered, we evict all MCUs from the cache that were not visible, i.e., MCUs whose visibility flag inside the cache was not set to true during the mark pass. The corresponding 16x16 blocks of decoded texels are returned to the pool for future allocations by the mark pass. The visibility flag of all retained MCUs is set to false, making them eligible for eviction in the next frame. 

\subsection{Multiple Textures and Mip Mapping}

Multiple textures are supported by combining the MCU ID with a texture ID into a single key for caching throughout the pipeline. Mip Mapping is supported by creating 7 downscaled versions of a texture and storing them as additional textures alongside the originals. The mip map level is then also included in the cache key. 

In our implementation we use 32 bit cache keys made of 16 bit for the MCU ID (supporting 4k textures with up to 256 x 256 MCUs), 13 bit for the texture ID (up to 8k unique textures), and 3 bit for the mip map level (up to 8 levels, including original).

\section{Implementation Details}
\label{sec:implementation_details}

This section provides additional details that are relevant for performance, overhead-reduction, and implementation. 

\subsection{Indexing Table}
\label{sec:indexing_table}

Random access to an MCU’s data requires an indexing table mapping MCU indices to their corresponding byte offsets. This auxiliary structure adds memory overhead compared to standard JPEG. To minimize the overhead, we store offsets as follows: the byte offset of every ninth MCU is stored absolutely in a 32-bit integer, while the offsets of the following eight MCUs are stored as 16-bit values relative to that absolute offset. 






\subsection{Warp-Level Parallel Huffman Decoding}

Decoding AC coefficients is a major bottleneck of the MCU decoding stage. Each AC is Huffman-coded with varying bit length, which requires us to sequentially decode up to $6 \cdot 63 = 378$ components per MCU in a single-threaded fashion. For each AC, we then need an additional inner loop in order to match the incoming bits with codes of the Huffman table. 

We optimize this inner loop as follows: First, we prefetch the next 16 bit of the compressed bit stream. We then use all 32 threads of the warp to compare to 32 Huffman codes at a time. Since each Huffman code may have a different bit length, the threads trim the prefetched bits as needed. During each iteration, all threads run a ballot to communicate whether they found a match. If a match is found, the winning thread broadcasts the results to the others.

\subsection{Memory Overhead}
\label{sec:overhead}

Enabling efficient random access requires adaptations of the JPEG format that increase the memory usage. This overhead consists of following components:

\begin{itemize}
    \item \textbf{Indexing table}: $\frac{32 + 16 \cdot 8}{9} = 17.777$ bit per MCU. The overhead of the indexing table is a result of storing the offset to every ninth MCU as an absolute 32 bit value, and the offsets to the following 8 MCUs using 16 bit values relative to the absolute offset.
    \item \textbf{DC Coefficients}: $3 \cdot 12 = 36$ bit per MCU. Each MCU has 6 DC coefficients -- 4 for luminance and 2 for chroma. JPEG encodes the DCs of the Y, Cr and Cb channels as the difference to the previous value of the respective channel, resulting in a sequential dependency to prior MCUs that we need to break to enable random access. We therefore store the DCs of Cr, Cb and the first DC of the Y channel in absolute terms as 12 bit values. The remaining 3 DCs of the Y component remain difference and Huffman coded, and thus do not add overhead. 
\end{itemize}

This gives us an upper bound of 53.777 bit per MCU, or $\frac{53.777}{16 \cdot 16} = 0.21$ bit per pixel. However, in practice the overhead is closer to 0.17 bit per pixel since we need to subtract the number of bits that the original JPEG allocated for the re-encoded DC components to obtain the actual overhead. 

\section{Evaluation} 

\newcommand{\figentry}[2]{%
    \begin{subfigure}[t]{0.195\linewidth}%
        \includegraphics[width=\textwidth]{#1}%
        \caption{#2}%
    \end{subfigure}%
}
\begin{figure*}[htp]
    \figentry{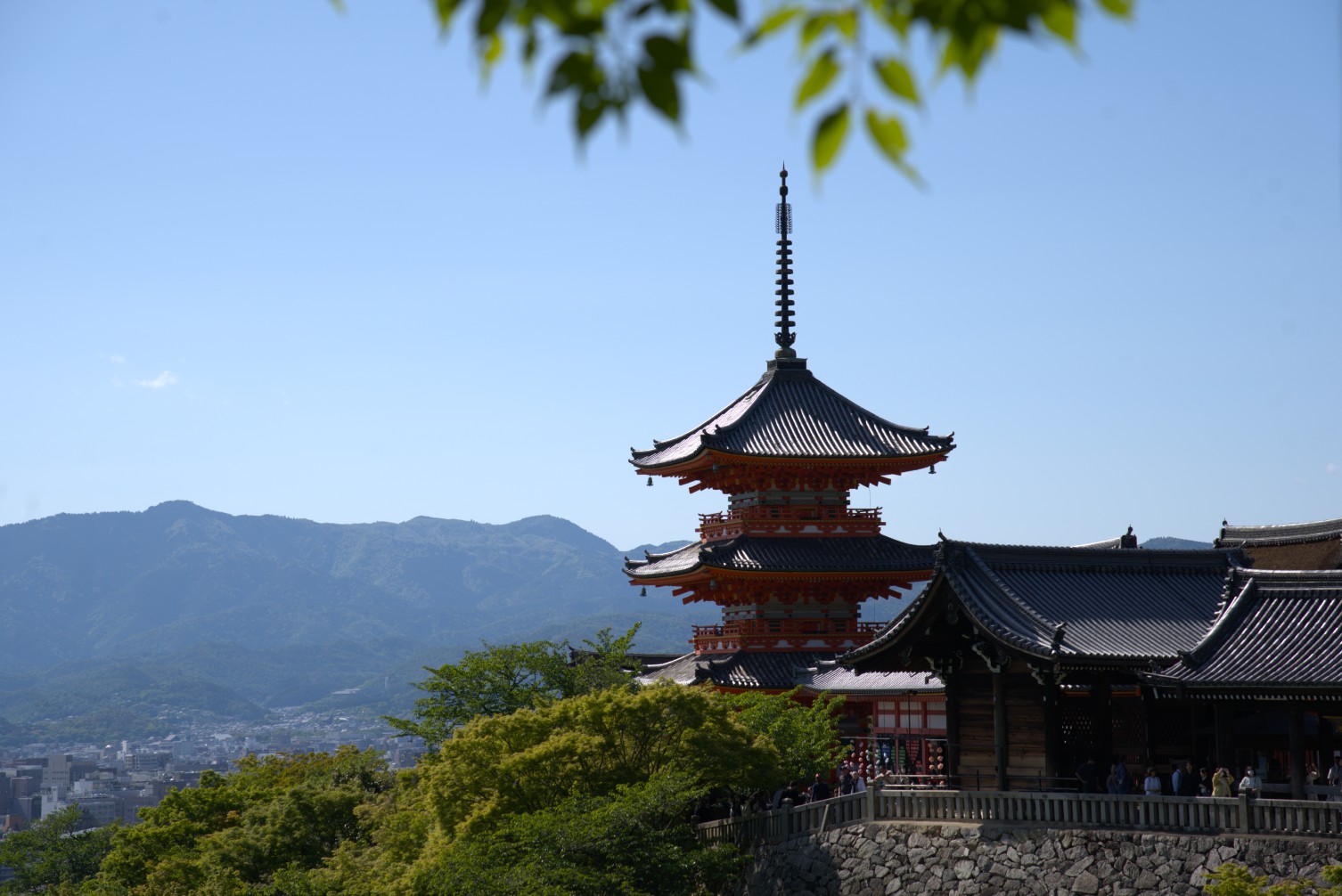}{Kiyomizu-Dera} \hfill
    \figentry{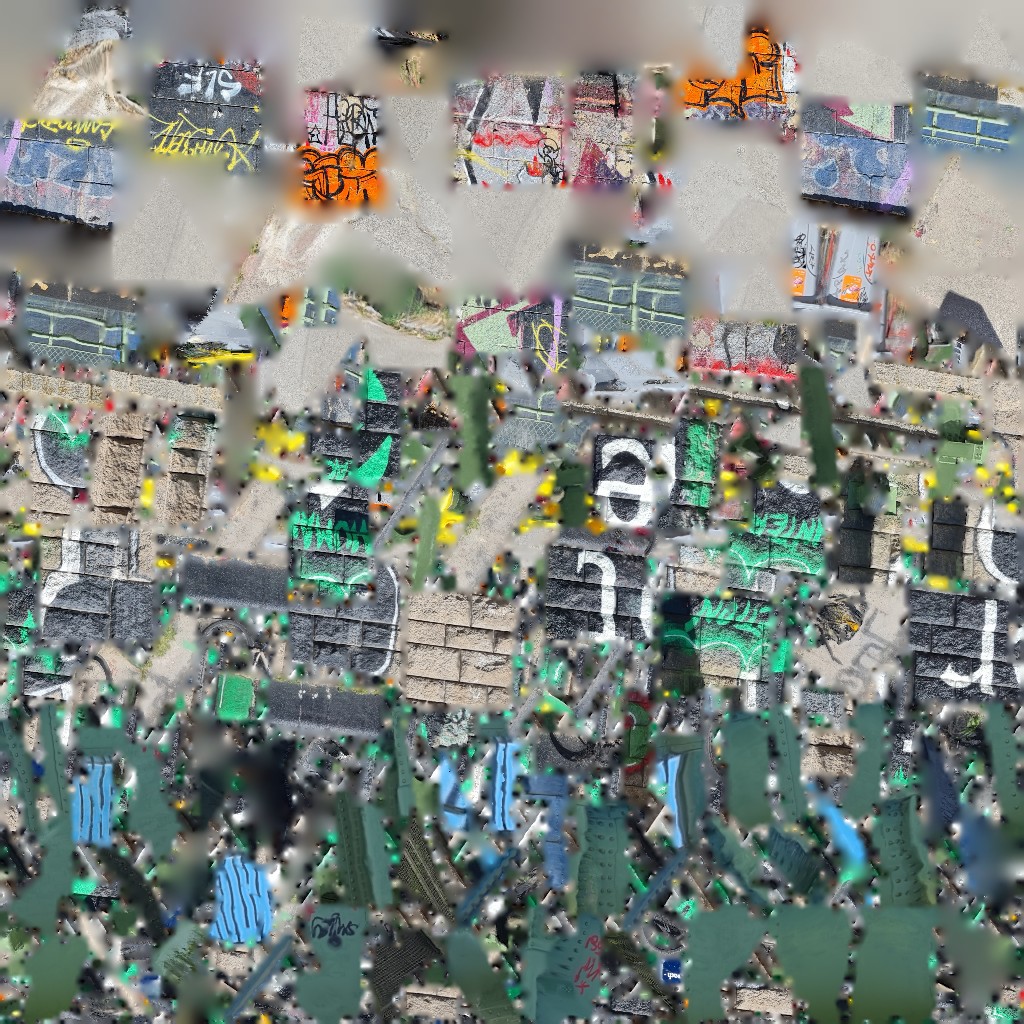}{Graffiti (tex3)} \hfill
    \figentry{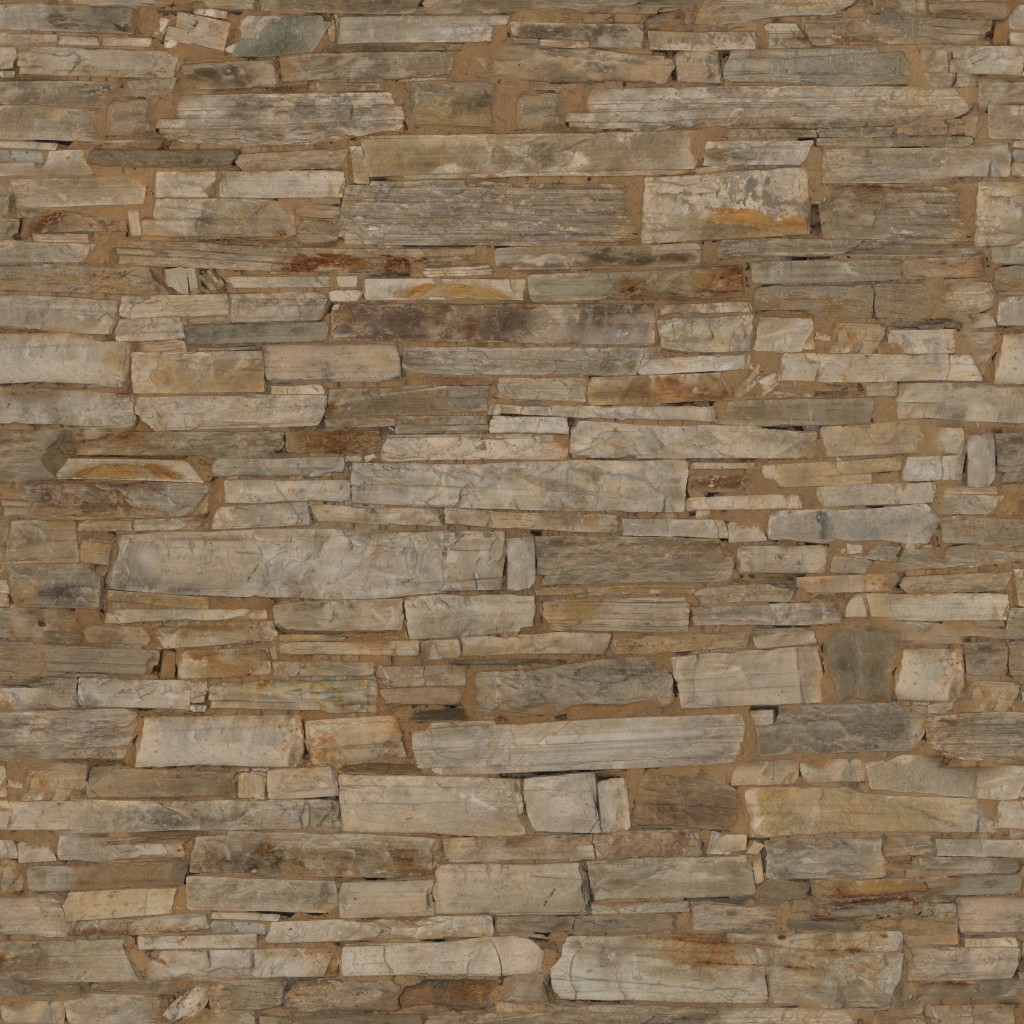}{Rustic Stone Wall 02} \hfill
    \figentry{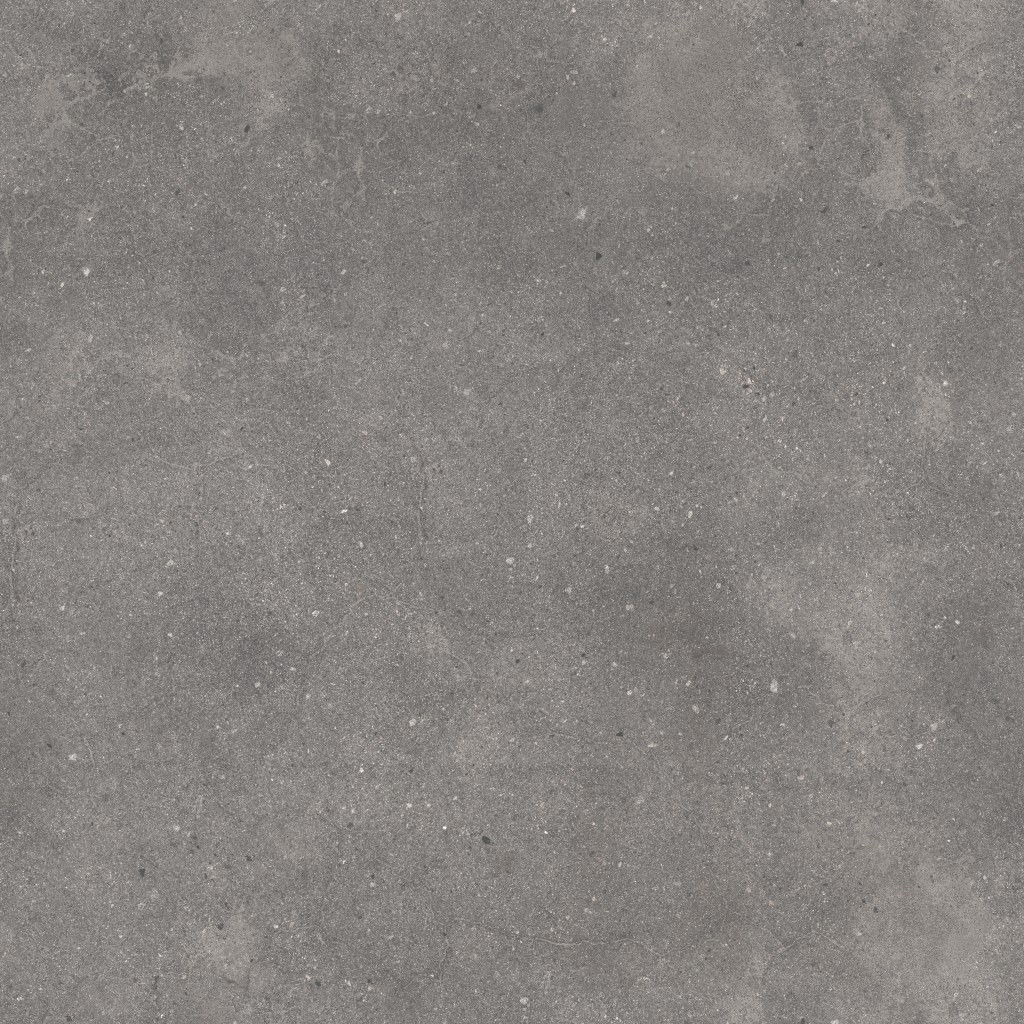}{Asphalt 04} \hfill
    \figentry{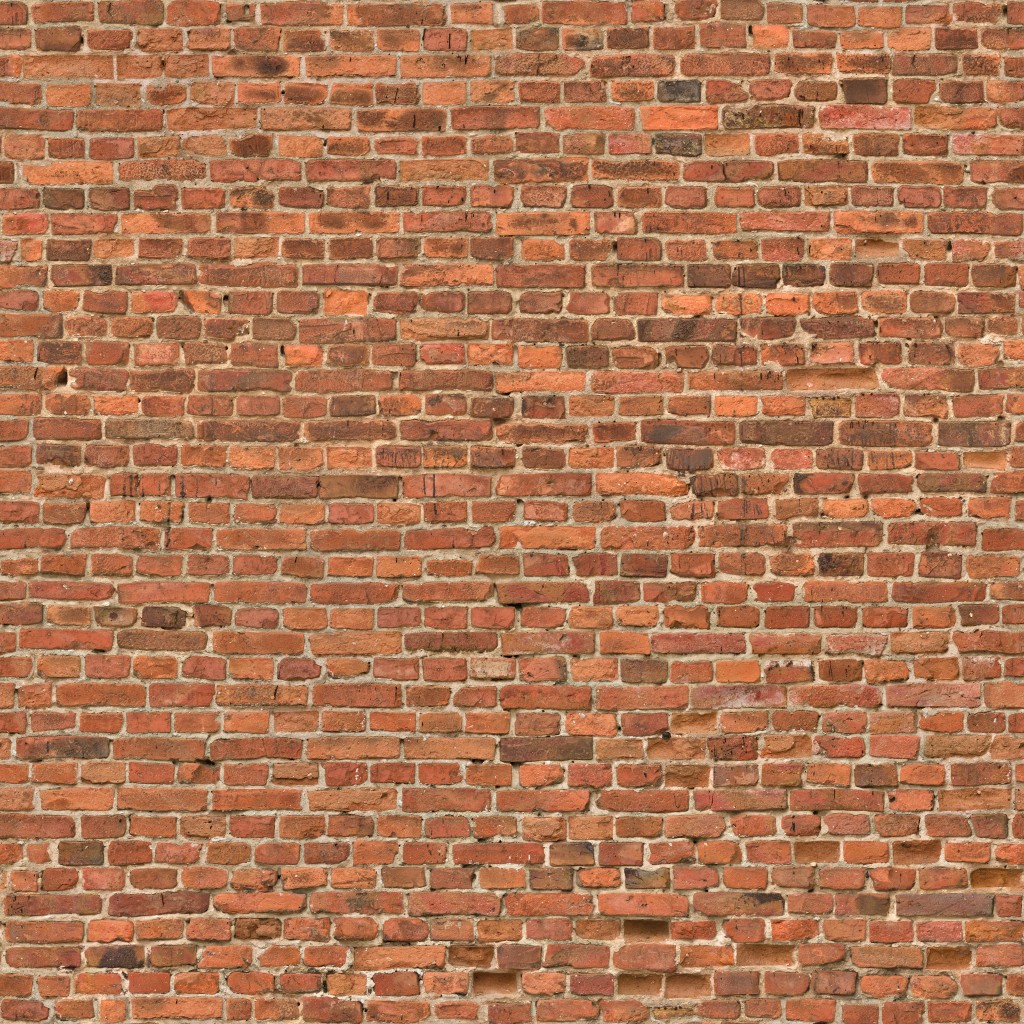}{Brick Wall 006} \hfill
    \figentry{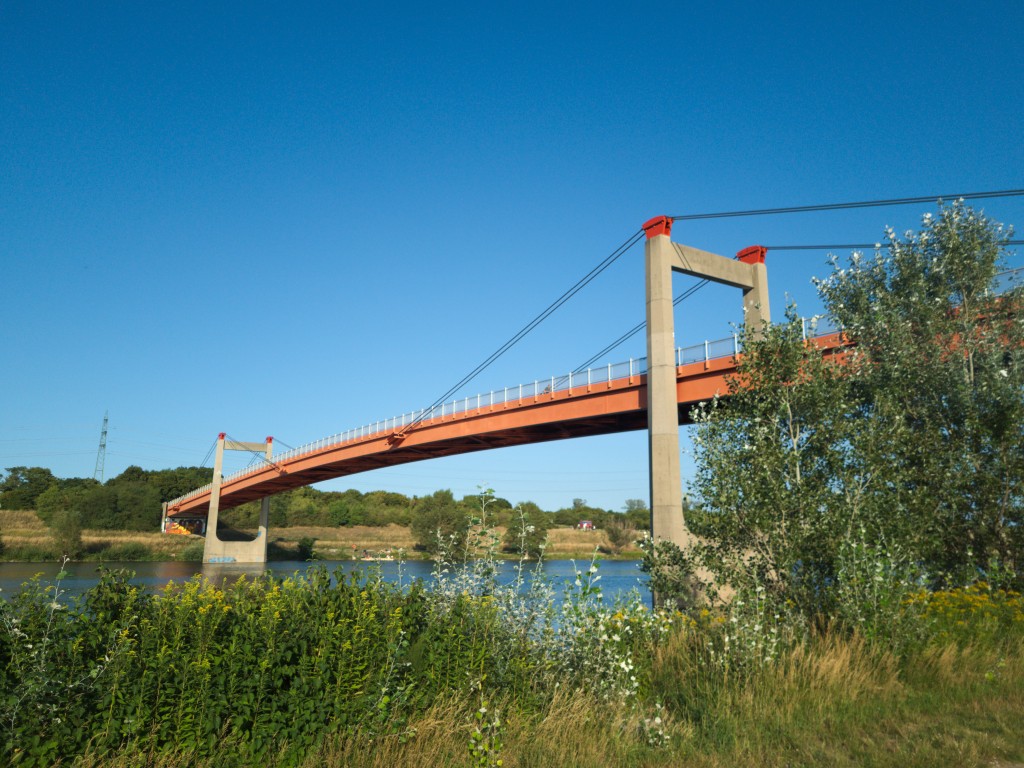}{Bridge} \hfill
    \figentry{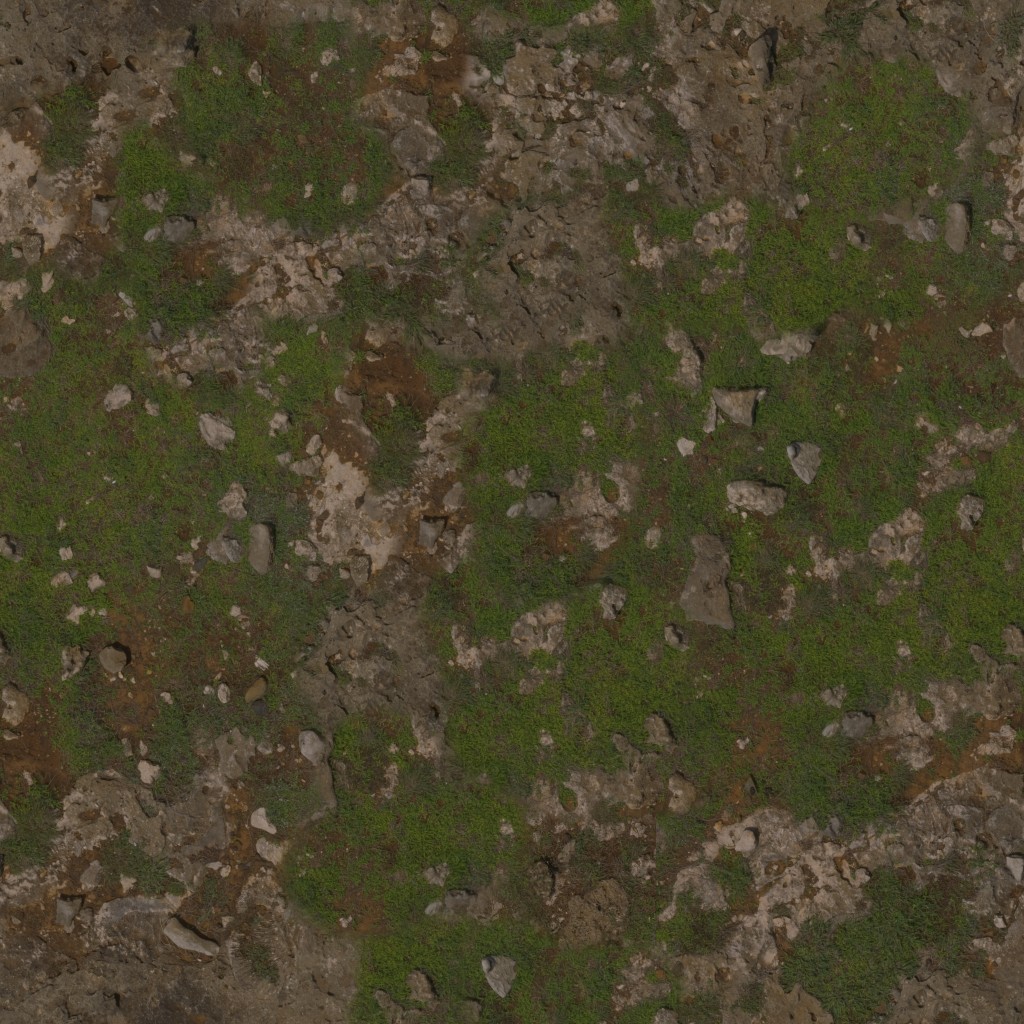}{Coast Sand} \hfill
    \figentry{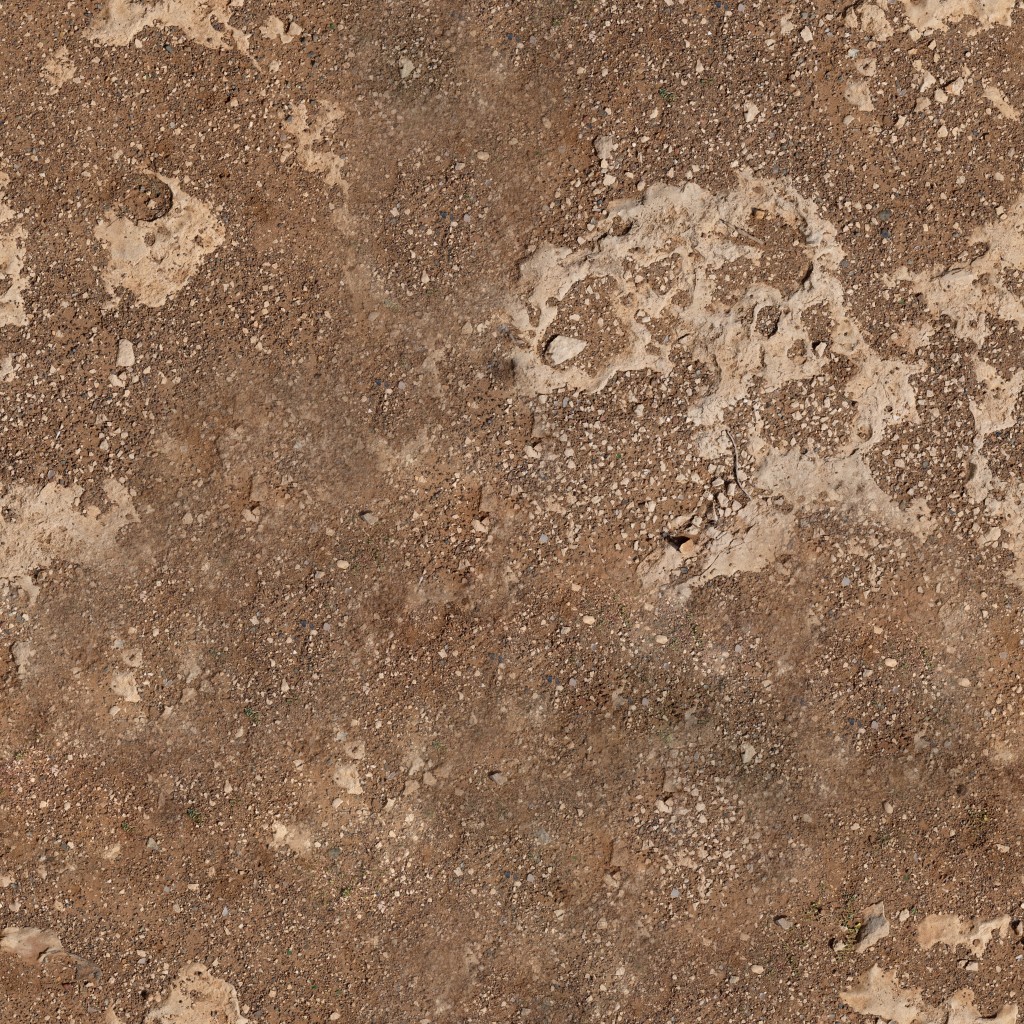}{Coral Gravel} \hfill
    \figentry{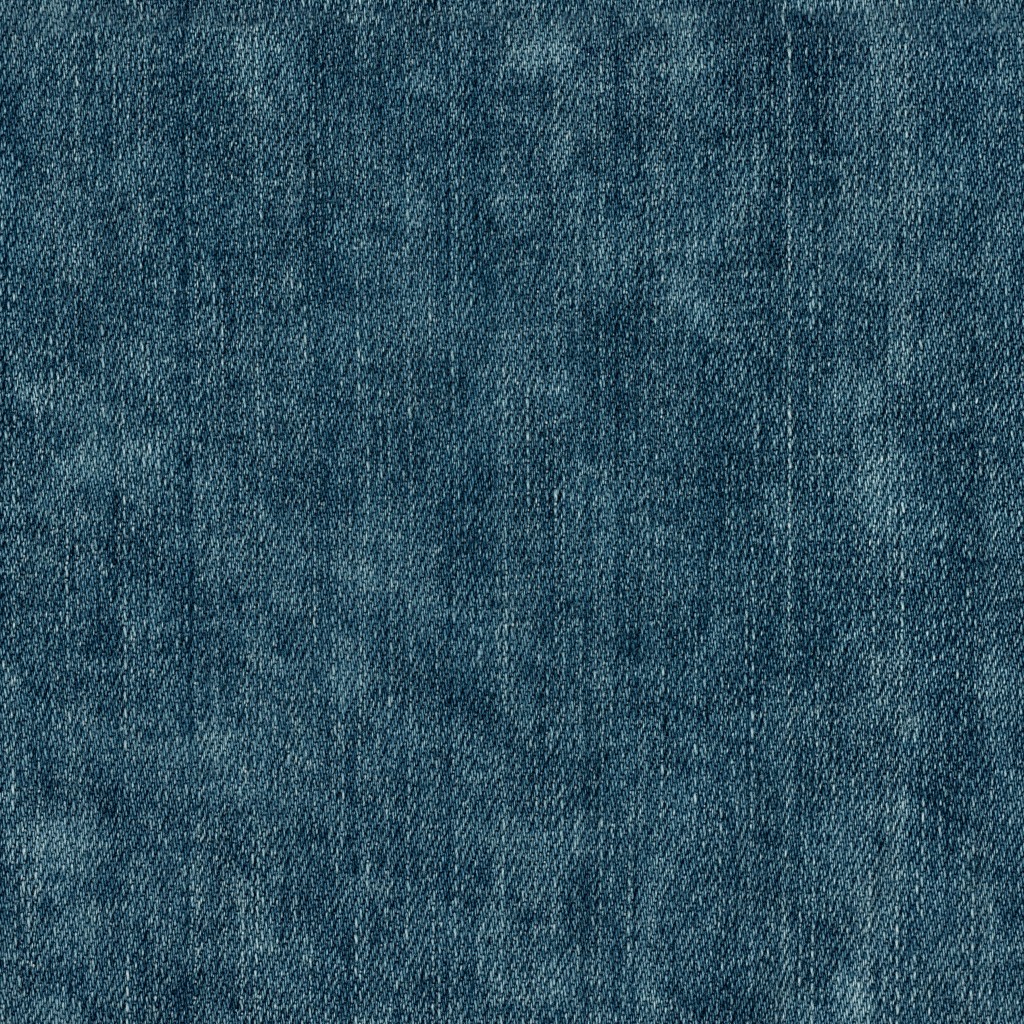}{Denim Fabric 02} \hfill
    \figentry{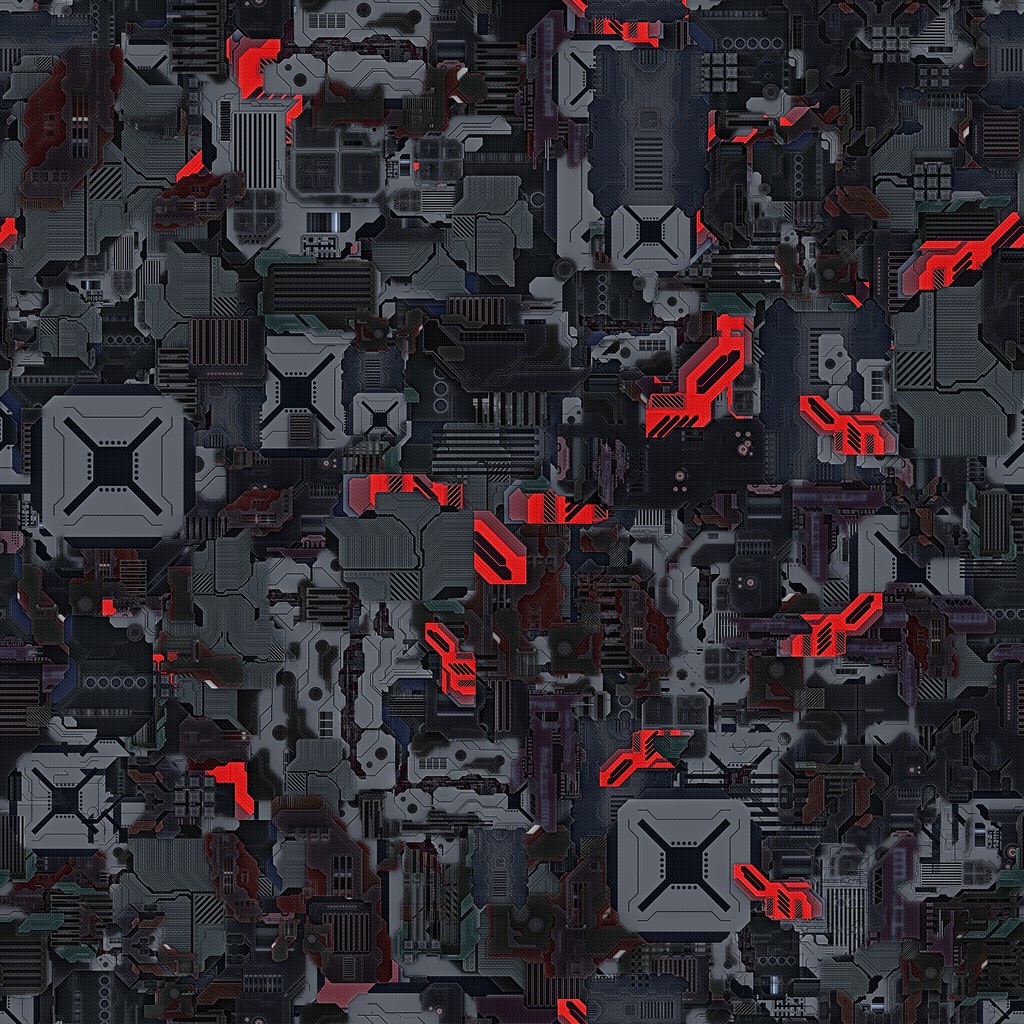}{Chip005} 
    \caption{Test textures: (a,f) Raw photos. (b) One of twenty 4k textures created by RealityScan during mesh construction. A part of the test scene "Graffiti". (rest) 4k textures from https://ambientcg.com/ and https://polyhaven.com/. }
    \label{fig:test_data}
\end{figure*}

To motivate the use of variable-rate compression, we evaluate both quality and bitrate across a range of image compression methods, including BC, ASTC, and JPEG, as well as newer formats such as AVIF and JPEG XL. We then evaluate their feasibility for real-time rendering by measuring the performance of our JPEG-based rendering pipeline in 3D scenes.

\subsection{Quality}

This section evaluates the trade-off between quality and size of various compression algorithms, allowing us to estimate the potential gains of using variable-rate compression in real-time-rendering. Although we currently only support JPEG with 4:2:0 chroma subsampling, we also compare with JPEG XL and AVIF as our work is intended to function as a stepping stone towards these more sophisticated formats. 

The test data (Figure~\ref{fig:test_data}) comprises a set of lossless compressed textures from https://ambientcg.com/ and https://polyhaven.com/; photos in raw format (to avoid a-priori compression artifacts); and a texture of a photogrammetry-based 3D model that was created with RealityScan. 

For encoding BC and ASTC, we use the NVIDIA texture tools exporter with compression quality set to "production". For JPEG, we use two encoders: libjpeg\_turbo and Google's JPEGLI. The former is a faster variation of libjpeg, while the latter also improves the quality of the constructed JPEG images. In both cases we use 4:2:0 chroma subsampling, since this is the configuration we support in our renderer. JPEG XL was encoded with Python's Pillow library. For AVIF, we use ffmpeg using libaom-av1.

The compression performance of Google's JPEG encoder "JPEGLI", plus 0.17 bit memory overhead, is representative for our method, and indicated by a dashed red line in Table~\ref{tab:quality_plots} and Table~\ref{tab:appendix_quality_plots}. 

The quality of the encoded images is evaluated with the metrics PSNR, FLIP~\cite{andersson2020flip}, SSIM~\cite{SSIM} and LPIPS~\cite{LPIPS}. PSNR scores are commonly used but also known to be flawed as they are not modeled after human perception. The others provide various solutions at scoring image quality based on perception, with FLIP being the most recent approach.

Across the scores shown in Table~\ref{tab:quality_plots} and Table~\ref{tab:appendix_quality_plots}, we can observe that BC1 typically takes the last place, followed by ASTC. AVIF dominates the PSNR, SSIM and LPIPS metrics, while JPEG XL leads with FLIP. JPEG lies in between, with Jpegli performing slightly better than libjpeg\_turbo. We can also observe several unintuitive results and disagreements between metrics. For example, LPIPS is the only metric that consistently scores Jpegli-encoded images worse than libjpeg\_turbo-encoded images; and SSIM most often scores ASTC as good as or sometimes better than JPEG. 

Overall, JPEG performs far better than BC, but surprisingly only modestly better than ASTC. JPEG XL and AVIF, on the other hand, are clearly superior, regularly encoding images in half the size for the same quality according to PSNR and FLIP, enabling sub-1bit-per-texel compression rates at sufficient quality. Table~\ref{tab:artifacts} and Table~\ref{tab:appendix_artifacts} provide impressions of the artifacts at a target bitrate of 0.89 bit per pixel.

\newcommand{\qcell}[2]{%
    \makecell{ \includegraphics[width=0.163\textwidth]{#1} \\ #2}
}

\newcommand{\qcellb}[2]{%
    \makecell{%
        \includegraphics[width=0.163\textwidth]{#1} \\%
        \begin{tabularx}{0.163\textwidth}{lXr}%
        #2 \\%
        \end{tabularx}%
    }%
}

\newcommand{\qcellc}[5]{%
    \noindent%
    \begin{tikzpicture}[baseline, inner sep=0pt, outer sep=0pt]%
    \node[anchor=south west,inner sep=0] (img) at (0,0) {\includegraphics[width=0.163\linewidth]{#1}};%
    \begin{scope}[x={(img.south east)},y={(img.north west)}] %
      \begingroup %
      \renewcommand{\textcolor}[2]{##2} %
      \node[anchor=north west, text=black!70!gray] at (0.025,0.97){\textcolor{black}{#2}};
      \node[anchor=north east, text=black!70!gray] at (0.990,0.97){\textcolor{black}{#3}};
      \node[anchor=south west, text=black!70!gray] at (0.025,0.01){\textcolor{black}{#4}};
      \node[anchor=south east, text=black!70!gray] at (0.990,0.01){\textcolor{black}{#5}};
      \endgroup
      \node[anchor=north west,text=white] at (0.02,0.98) {#2};  %
      \node[anchor=north east,text=white] at (0.98,0.98) {#3};  %
      \node[anchor=south west,text=white] at (0.02,0.02) {#4};  %
      \node[anchor=south east,text=white] at (0.98,0.02) {#5};  %
    \end{scope} %
  \end{tikzpicture}%
}

\newcommand{\plotrow}[4]{%
    \includegraphics[width=\IMGWIDTH\textwidth, trim=0cm #3 0cm 1mm, clip]{#1/plot_psnr.png} & %
    \includegraphics[width=\IMGWIDTH\textwidth, trim=0cm #3 0cm 1mm, clip]{#1/plot_flip.png} &%
    \includegraphics[width=\IMGWIDTH\textwidth, trim=0cm #3 0cm 1mm, clip]{#1/plot_lpips.png} &%
    \begin{tikzpicture}%
        \node[inner sep=0pt] (bg) {\includegraphics[width=\IMGWIDTH\textwidth, trim=0cm #3 0cm 1mm, clip]{#1/plot_ssim.png}};%
        \node[inner sep=0pt, anchor=south east, yshift=#4, xshift=-3pt] at (bg.south east) {\includegraphics[width=0.1\textwidth]{#2}};%
    \end{tikzpicture} \\%
}

\definecolor{winnerColor}{RGB}{50,200,50}
\definecolor{loserColor}{RGB}{200,50,50}
\newcommand{\winner}[1]{\textcolor{winnerColor}{\textbf{#1}}}
\newcommand{\loser}[1]{\textcolor{loserColor}{\textbf{#1}}}

{
\setlength{\tabcolsep}{1pt}
\newcommand{\IMGWIDTH}{0.249}

\begin{table*}[]
\caption{Artifacts of various compression algorithms at a target bitrate of 0.89 bit per pixel (bpp), or the quality level that comes closest. BC1 for comparison (always 4bpp). Each cell shows quality and bpp in the top row, and FLIP and PSNR in the bottom row. Best and worst FLIP and PSNR values in a row are highlighted, with the exception of BC1 which is not considered due to its high memory usage.
}

\begin{tabular*}{\textwidth}{cc||cccc}
  \textbf{Reference}  & \textbf{BC1} & \textbf{ASTC 12x12} & \textbf{JPEG} & \textbf{JPEG XL} & \textbf{AVIF} \\ 
  \qcellc{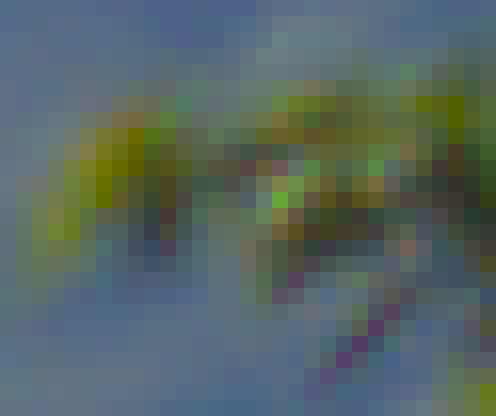}                  {Kiyomizu-Dera}{}{}{} & 
  \qcellc{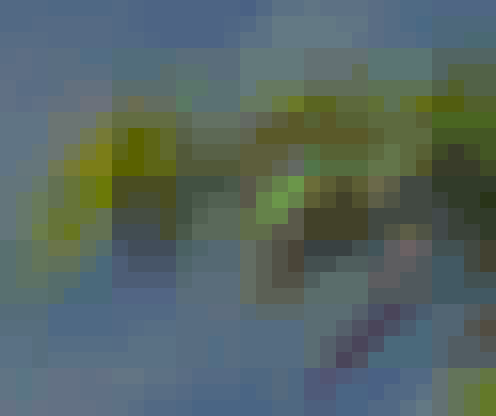}                            {   }{4.00bpp} {0.031}          {38.4}   & 
  \qcellc{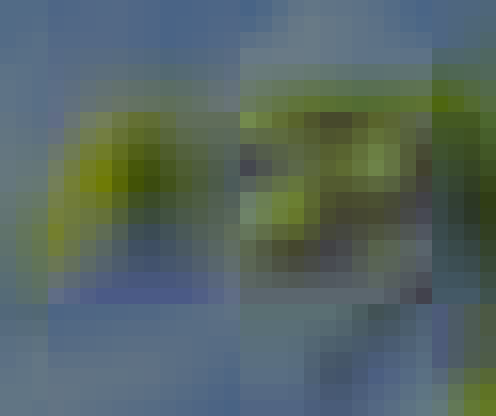}                 {   }{0.89bpp} {\loser{0.038}}  {\loser{36.5}}  & 
  \qcellc{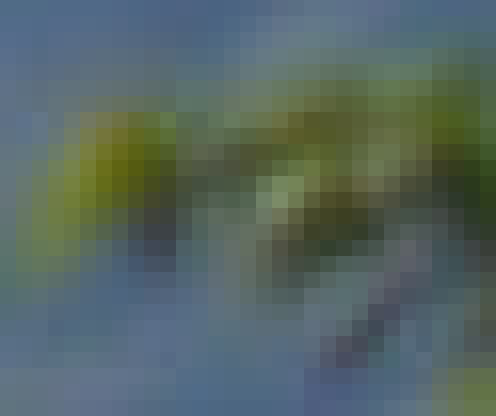}                      {Q90}{0.64bpp} {0.037}          {37.4}  & 
  \qcellc{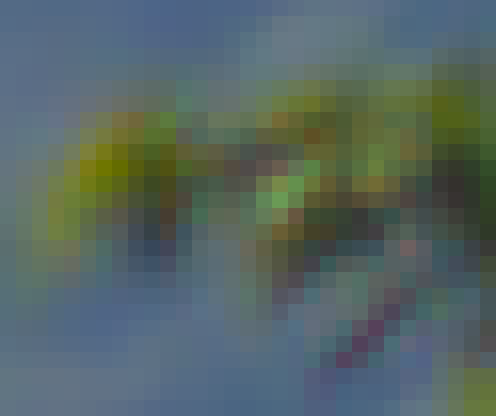}                      {Q90}{0.74bpp} {\winner{0.027}} {39.4}  & 
  \qcellc{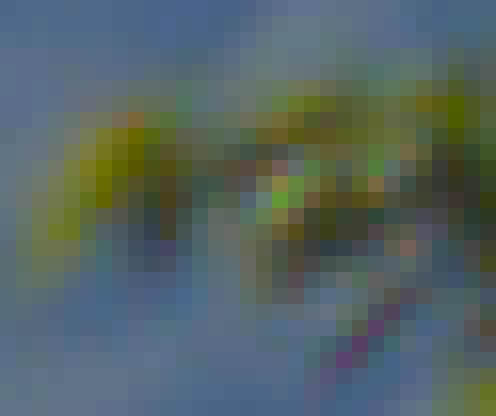}                        {Q85}{0.84bpp} {0.034}          {\winner{42.8}}   \\
  \qcellc{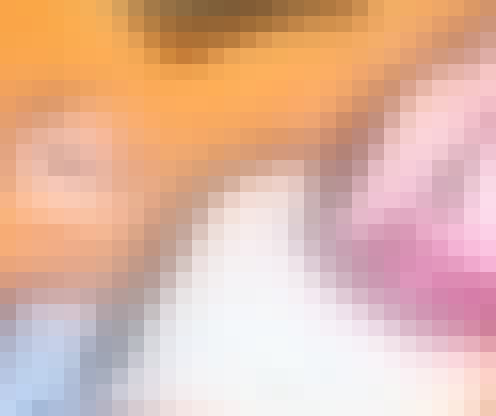}                 {Graffiti (Tex3)}{}{}{} &
  \qcellc{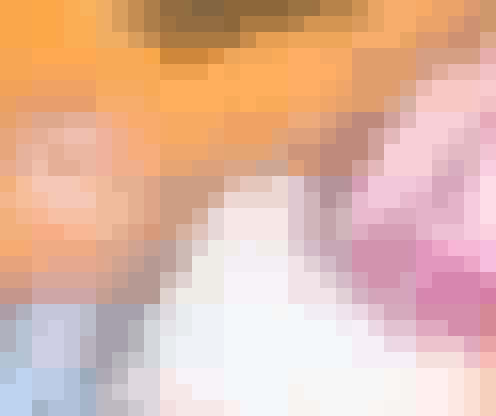}                           {   } {4.00bpp} {0.033}          {38.8}  &
  \qcellc{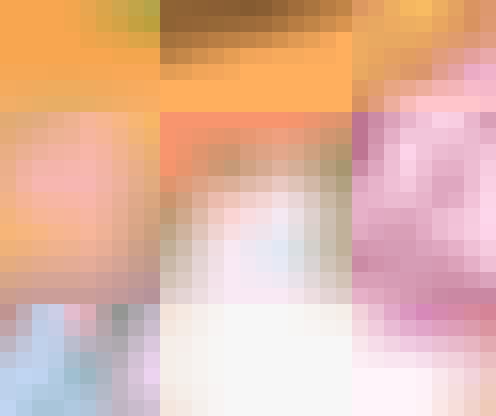}                {   } {0.89bpp} {\loser{0.051}}  {\loser{33.2}}   &
  \qcellc{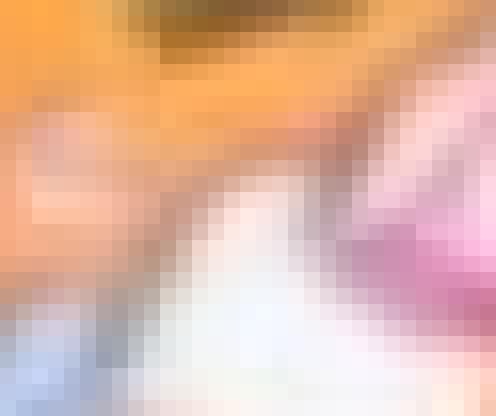}                     {Q80} {0.90bpp} {0.042}          {37.3} &
  \qcellc{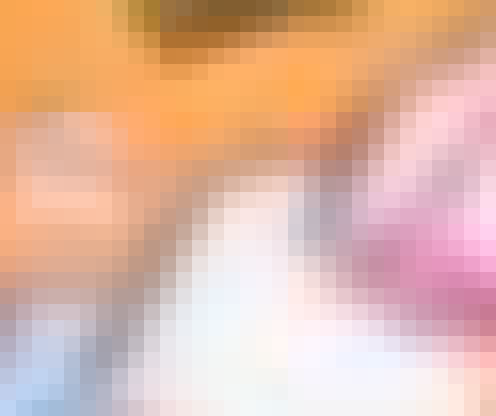}                     {Q80} {0.77bpp} {\winner{0.035}} {37.9}    &
  \qcellc{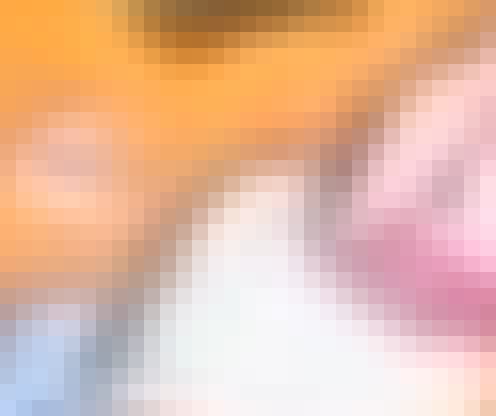}                       {Q75} {0.89bpp} {0.038}          {\winner{41.7}}   \\
  \qcellc{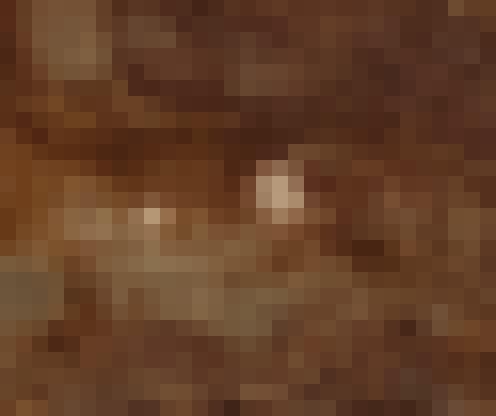}{Rustic Stone Wall 02}{}{}{} &
  \qcellc{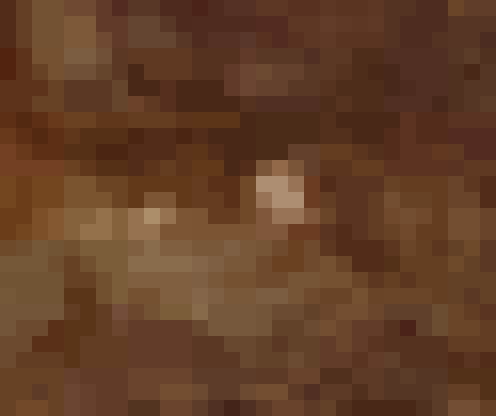}            {   }{4.00bpp}{0.032}{41.6} &
  \qcellc{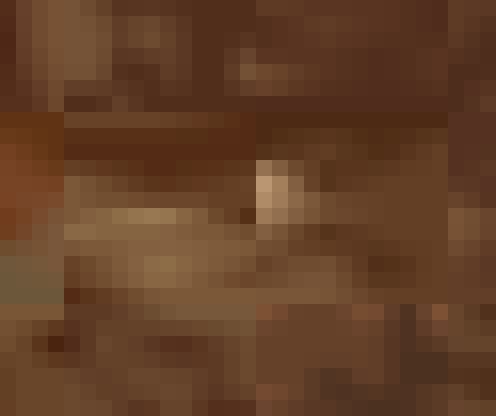} {   }{0.89bpp}{0.048}{\loser{37.4}} &
  \qcellc{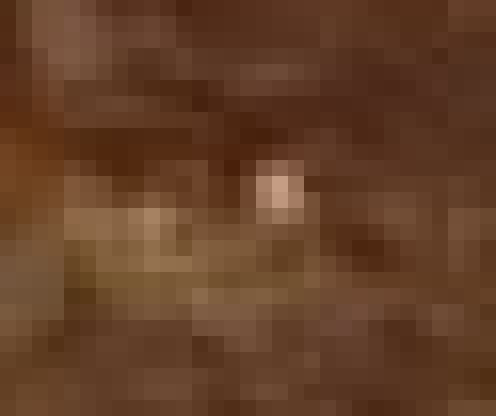}      {Q80}{0.77bpp}{\loser{0.047}}{37.7} &
  \qcellc{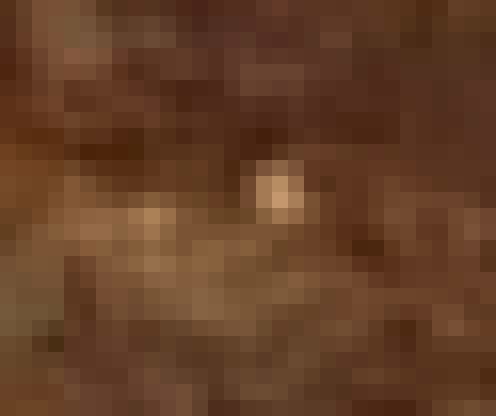}      {Q80}{0.74bpp}{\winner{0.039}}{38.7} &
  \qcellc{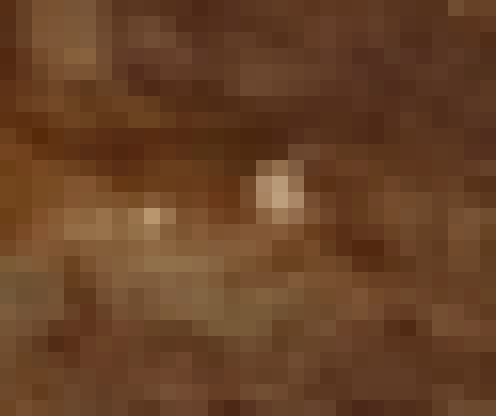}        {Q80}{0.96bpp}{\winner{0.039}}{\winner{40.4}} \\
\end{tabular*}
\label{tab:artifacts}
\end{table*}

\begin{table*}[]
\caption{Quality metric plots for the test data sets in Table~\ref{tab:artifacts}. "JPEGLI + 0.17 bit" is representative of our work, which adapts JPEG but adds about 0.17 bit per pixel to enable random access.}
\begin{tabular*}{\textwidth}{cccc}
    PSNR↑ & FLIP↓ & LPIPS↓ & SSIM↑ \\ 
    \plotrow{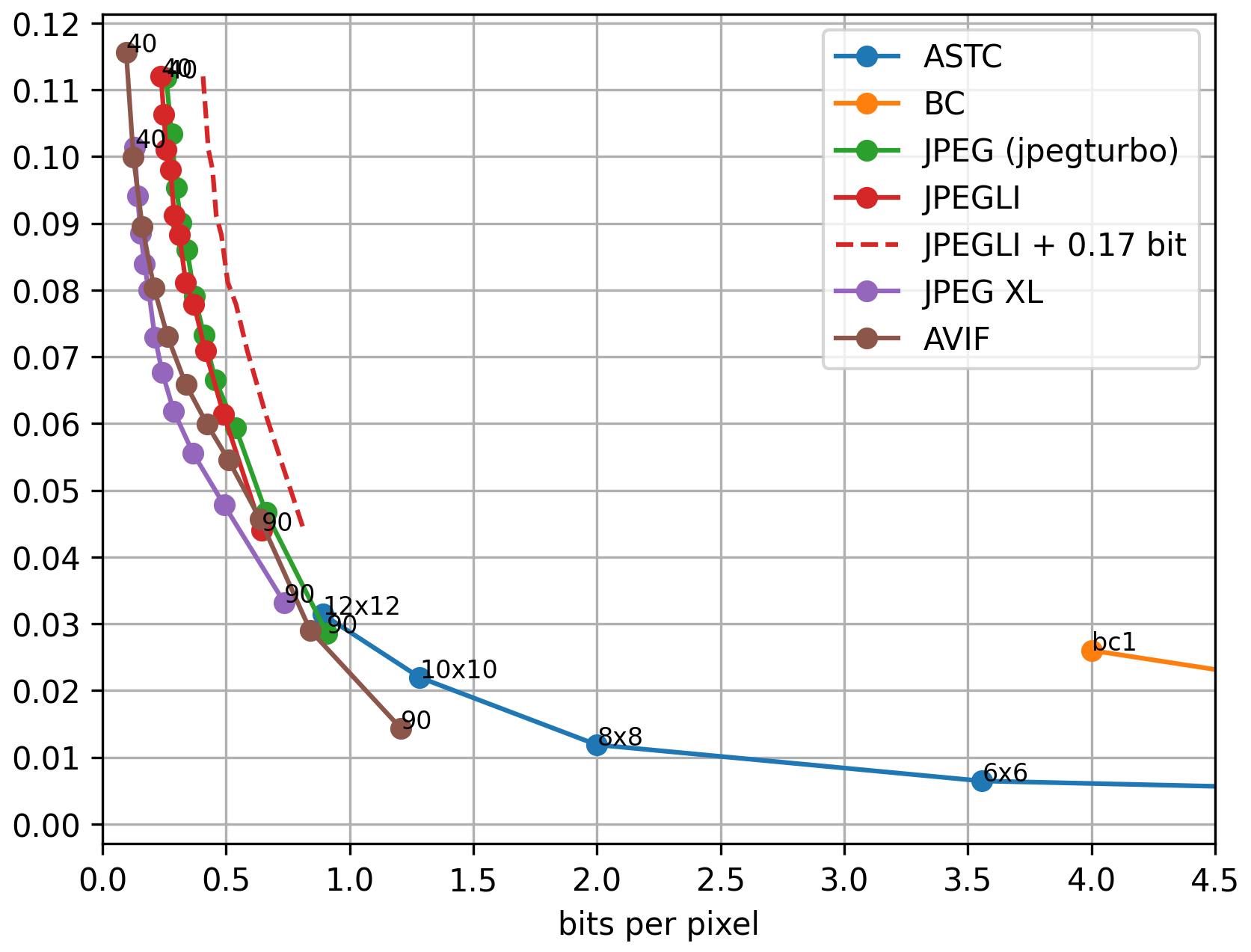}{images/qualitative/kiyomizudera/kiyomizu-dera.jpg}{34px}{0.3pt}
    \plotrow{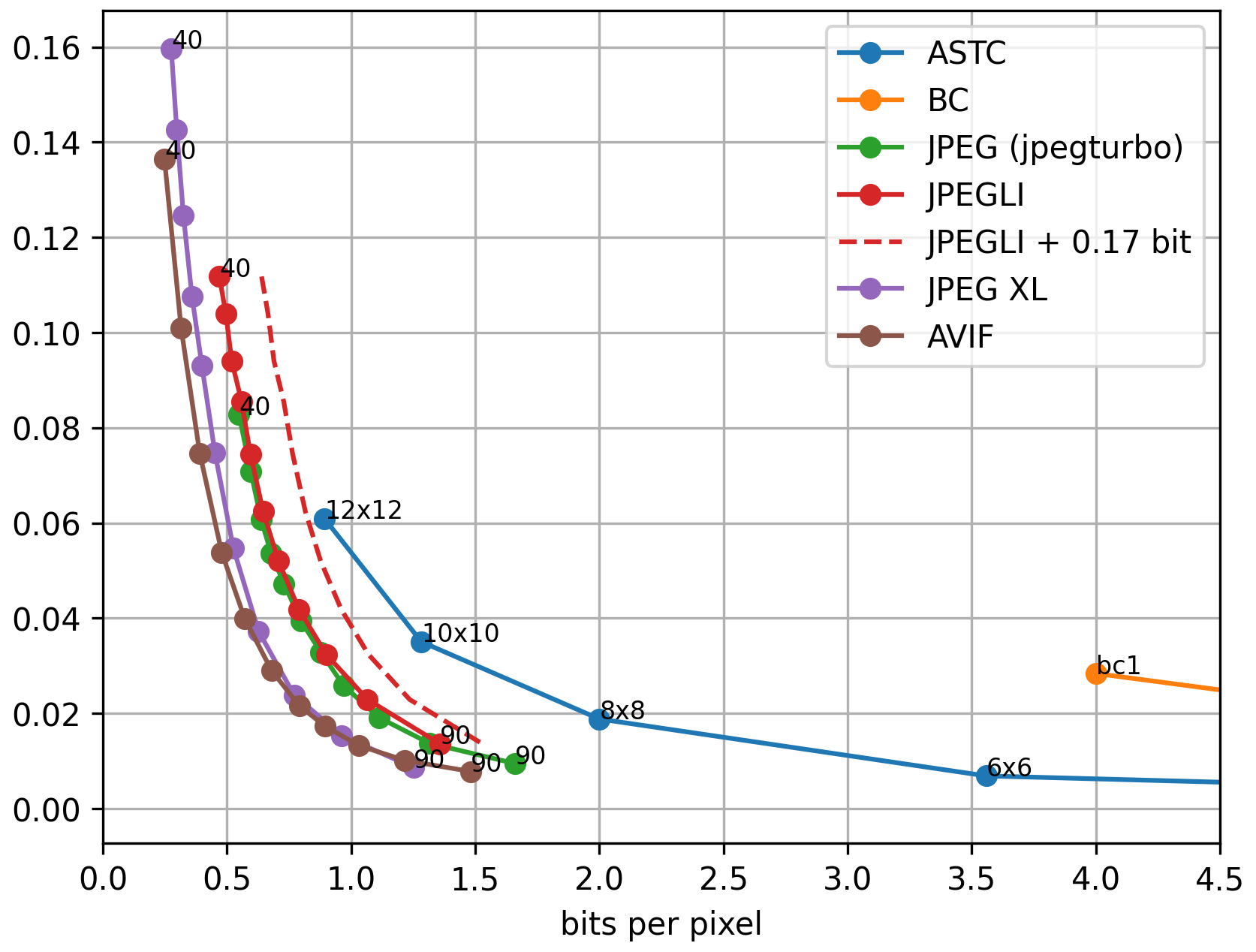}{images/qualitative/graffiti_tex3/graffiti_tex3.jpg}{34px}{0.3pt}
    \plotrow{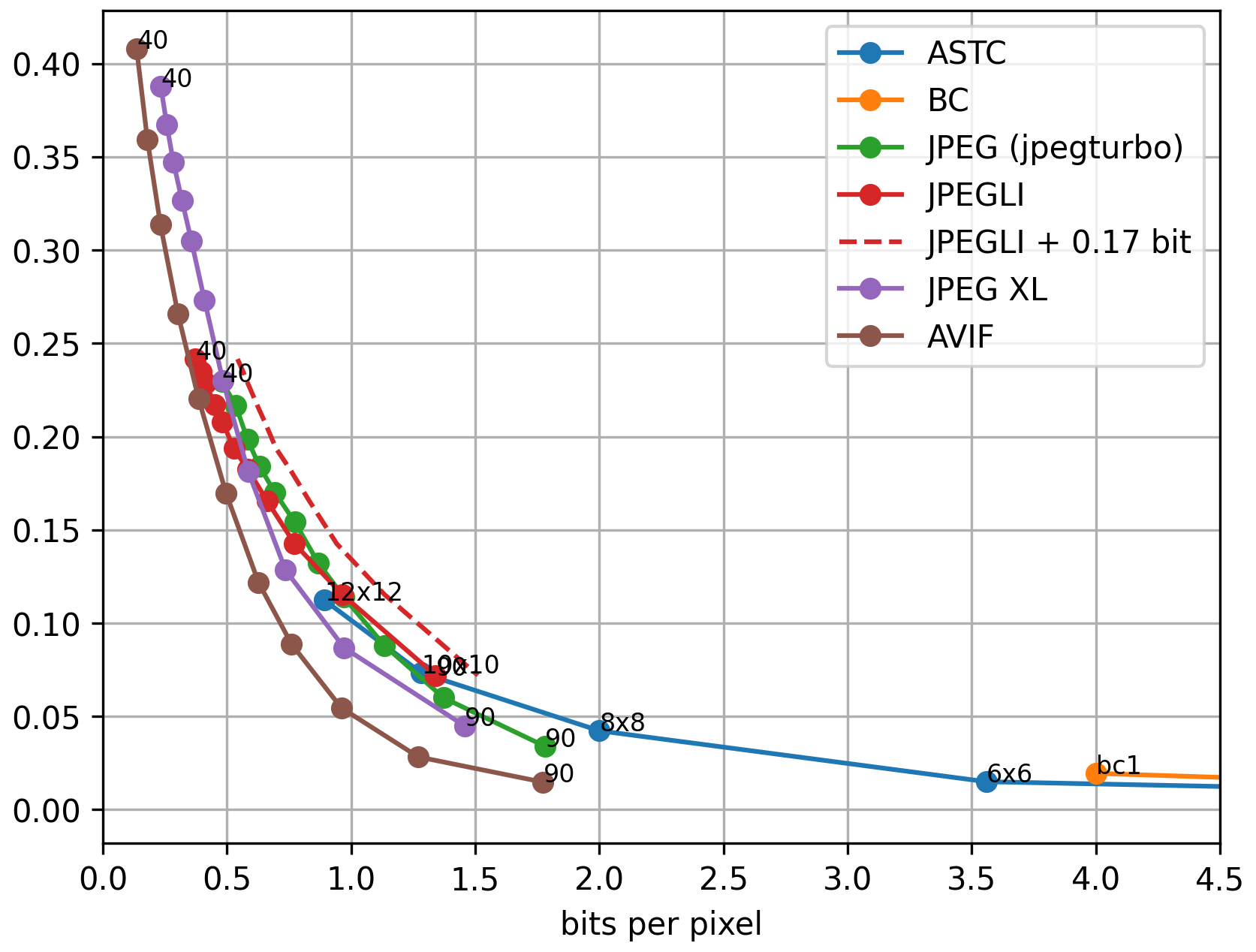}{images/qualitative/rustic_stone_wall_02_diff_4k/rustic_stone_wall_02_diff_4k.jpg}{0px}{10.83pt}
\end{tabular*}
\label{tab:quality_plots}
\end{table*}
}


\subsection{Performance}

Performance is evaluated in a CUDA- and OpenGL-based rendering engine. OpenGL renders a G-Buffer comprising uv-coordinates, texture-IDs and mip map levels. CUDA decodes the required JPEG blocks and subsequently converts the G-Buffer into a textured rendering. OpenGL draw call durations are computed with timestamp queries, and CUDA kernel launch durations with CUDA events. All durations are computed as the median over 60 frames. Two test scenes (Table~\ref{tab:test_scenes}) were evaluated: The popular "Crytek Sponza"~\cite{McGuire2017Data}, and a custom 3D scan "Graffiti" that consists of high-resolution, non-repeating textures.

{
\setlength{\tabcolsep}{0pt}
\begin{table}[H]
\centering
\caption{Test Scenes. }
\begin{tabular}{cc}
 Sponza & Graffiti \\
 \includegraphics[width=0.495\columnwidth]{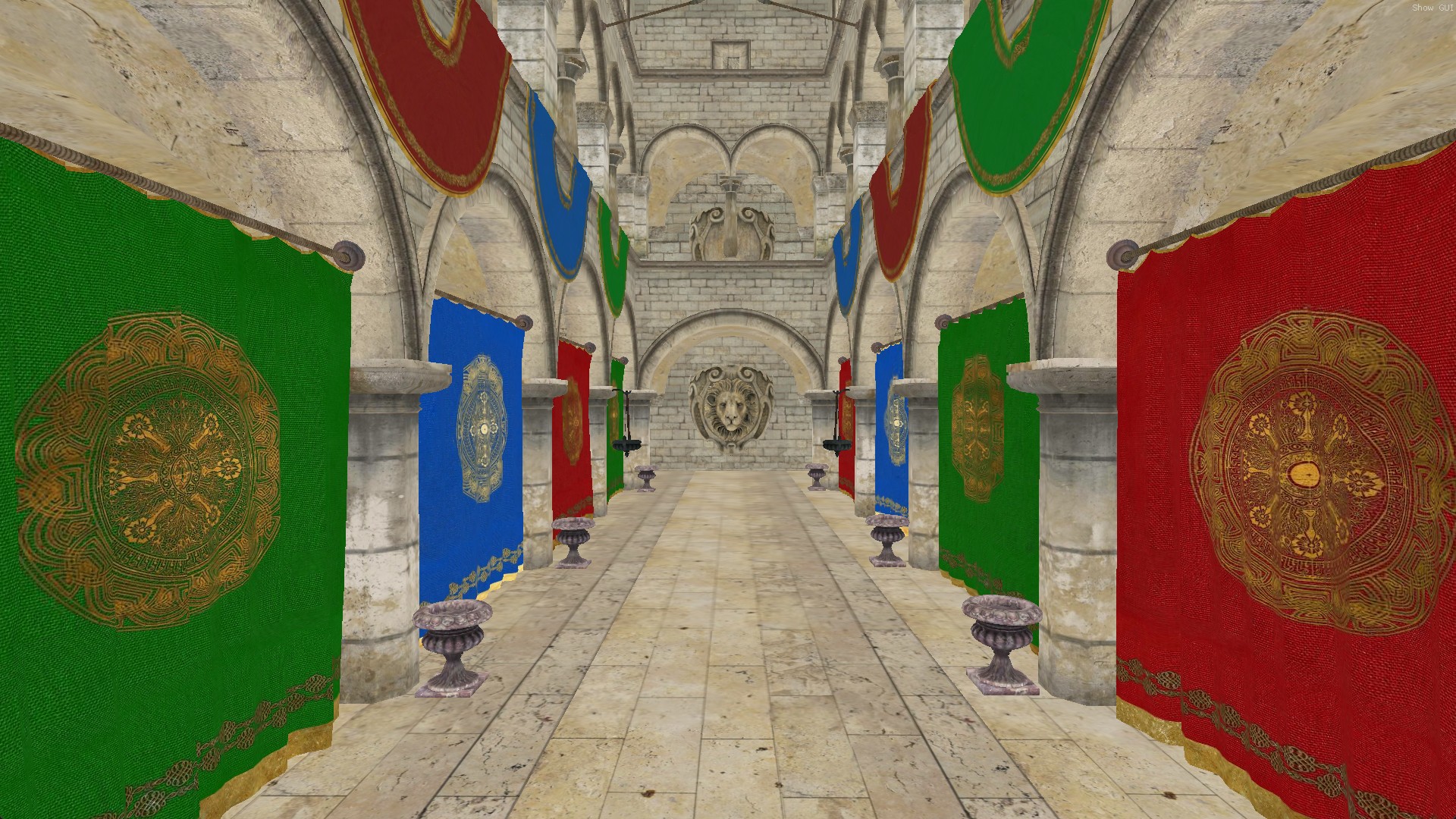} &  
 \includegraphics[width=0.495\columnwidth]{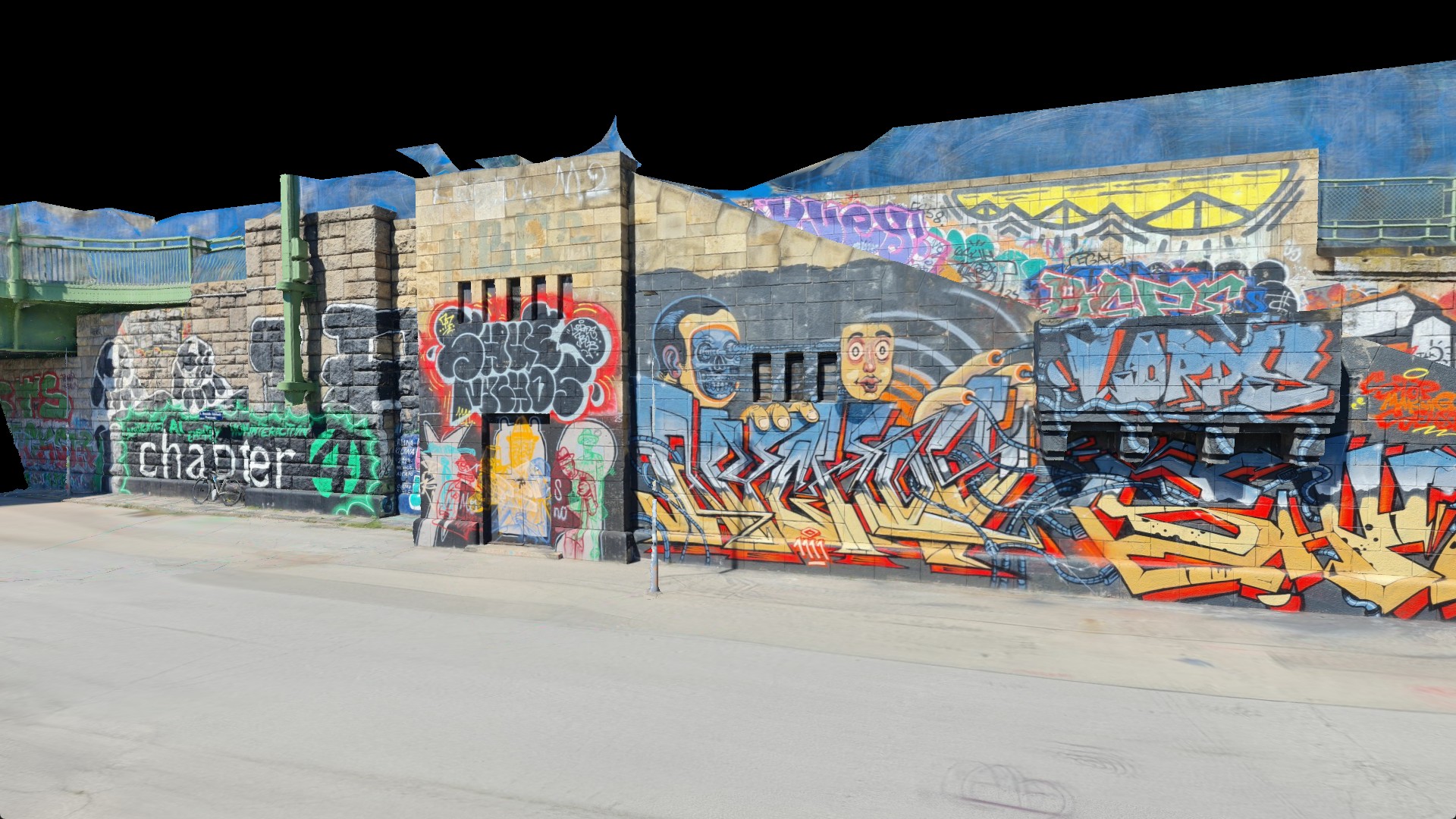} \\
 262k Triangles & 966k Triangles \\
 92 x 1k textures (96M texel) & 20 x 4k textures (335M texel) \\
 Repeated UVs (e.g. floor) & Non-Repetitive Texture Mapping
\end{tabular}
\label{tab:test_scenes}
\end{table}
}

Performance was measured on a test system with an AMD Ryzen 9 7950X CPU, an RTX 4090 GPU, and a Valve Index HMD. Framebuffer sizes are 1920 x 1080 (2MP) on desktop, and 2468 x 2740 per eye in VR (2 x 6.8MP). 

\subsubsection{Baseline Rendering}
Table~\ref{tab:peformance} reports the rendering performance across two test scenes and multiple JPEG quality levels with texture block caching disabled and linear interpolation enabled. 
Two key observations can be made: (1) higher JPEG quality levels significantly increase decoding time, and (2) mip mapping reduces decoding time to a fraction. The quality-dependent slowdown is explained by the number of AC coefficients: a higher quality implies more coefficients to decode. In contrast, mip mapping boosts performance by reducing the number of MCUs that are visible, as numerous small ones are replaced by fewer larger ones that better match their projected size. It also provides substantial performance improvements to the resolve stage as it increases the likelihood that adjacent pixels fetch texels from the same MCU. 

{
\setlength{\tabcolsep}{3pt}
\begin{table}[]
\caption{Baseline performance without caching (milliseconds). The geometry pass renders a G-Buffer in OpenGL. Mark, decode and resolve are CUDA kernels. Mip Mapping (MM) greatly reduces the amount of MCUs that need decoding.}
\label{tab:peformance}
\begin{tabular}{l|c|rrrr|r}
               & MM            & geometry & mark & decode & resolve & MCUs   \\
\hline
Sponza Q50     &               & 0.05     & 0.06 & 0.80   & 0.09   & 81k    \\
               & \checkmark    & 0.05     & 0.03 & 0.23   & 0.05    & 12k    \\
\hline
Sponza Q70     &               & 0.05     & 0.06 & 1.08   & 0.09   & 81k    \\
               & \checkmark    & 0.05     & 0.03 & 0.26   & 0.05    & 12k    \\
\hline
Sponza Q90     &               & 0.05     & 0.06 & 1.90   & 0.09   & 81k    \\
               & \checkmark    & 0.05     & 0.03 & 0.31   & 0.05    & 12k    \\
\hline
Graffiti Q80   &               & 0.12     & 0.12 & 2.63   & 0.19   & 224k   \\
               & \checkmark    & 0.12     & 0.04 & 0.46   & 0.05    & 22k    \\
\end{tabular}%
\end{table}
}

\subsubsection{Caching}
\label{sec:caching}
Caching must be evaluated under motion, since rendering the same viewpoint twice reduces the number of MCUs to decode to zero. To simulate fast-paced motion, we rotate the camera by 6 degrees to the right each frame, for a full rotation every 60 frames. Since we are mainly interested in the worst-performing frame to ensure we do not suffer from stutters, we compute and report the max-of-medians: (1) We repeat the 60 rotations 100 times, i.e., we sample a total of 6000 frames, (2) We compute the median of the 100 samples of a view direction, for all view directions, (3) we compute the maximum over the 60 median values to obtain the median of the worst-performing viewpoint. The reason for not using max by itself is that frame times and kernel launches are subject to fluctuations (hardware, scheduling, OS interrupts, ...), which is why the median is typically used. 

The results of the JPEG rendering pipeline in motion, with linear interpolation enabled, are shown in Table~\ref{tab:peformance_motion}. From this we can conclude that the JPEG pipeline adds $\leq 0.3$ ms to the frame on an RTX 4090, rendering into a framebuffer with a size of 1920 x 1080 pixels.

\begin{table}[]
\caption{JPEG pipeline (mark+decode+resolve) duration under motion. Reporting the max-of-medians, i.e., the median of multiple repetitions of the worst-performing viewpoint.}
\label{tab:peformance_motion}
\begin{tabularx}{\columnwidth}{lrrr}
               & Baseline & w. Mip Map & w. Mip Map \& Cache \\
\hline
Sponza Q50     &  1.01 ms &  0.35 ms   &    0.19 ms \\
Sponza Q70     &  1.27 ms &  0.36 ms   &    0.20 ms \\
Sponza Q90     &  2.14 ms &  0.43 ms   &    0.21 ms \\
Graffiti Q80   &  3.90 ms &  0.59 ms   &    0.30 ms \\
\end{tabularx}%
\end{table}

\subsubsection{Linear Interpolation}

Linear Interpolation between magnified texels has only a minor impact on overall rendering performance. In Sponza, the resolve kernel requires 25 µs with nearest-neighbor interpolation and 46 µs with linear interpolation. Most of this overhead arises when the interpolated texels lie in different MCUs, since each texel fetch triggers a hash map lookup into the texture block cache. When they reside in the same MCU, the result can be shared, reducing the cost. If we clamp texel fetches during linear interpolation to a single MCU, the cost is reduced to 32 µs, but this leads to artifacts along the borders of a 16x16 block of texels. 

\subsubsection{Virtual Reality}

\begin{figure}[]
    \centering
    \begin{subfigure}[t]{\linewidth}
        \includegraphics[width=\linewidth]{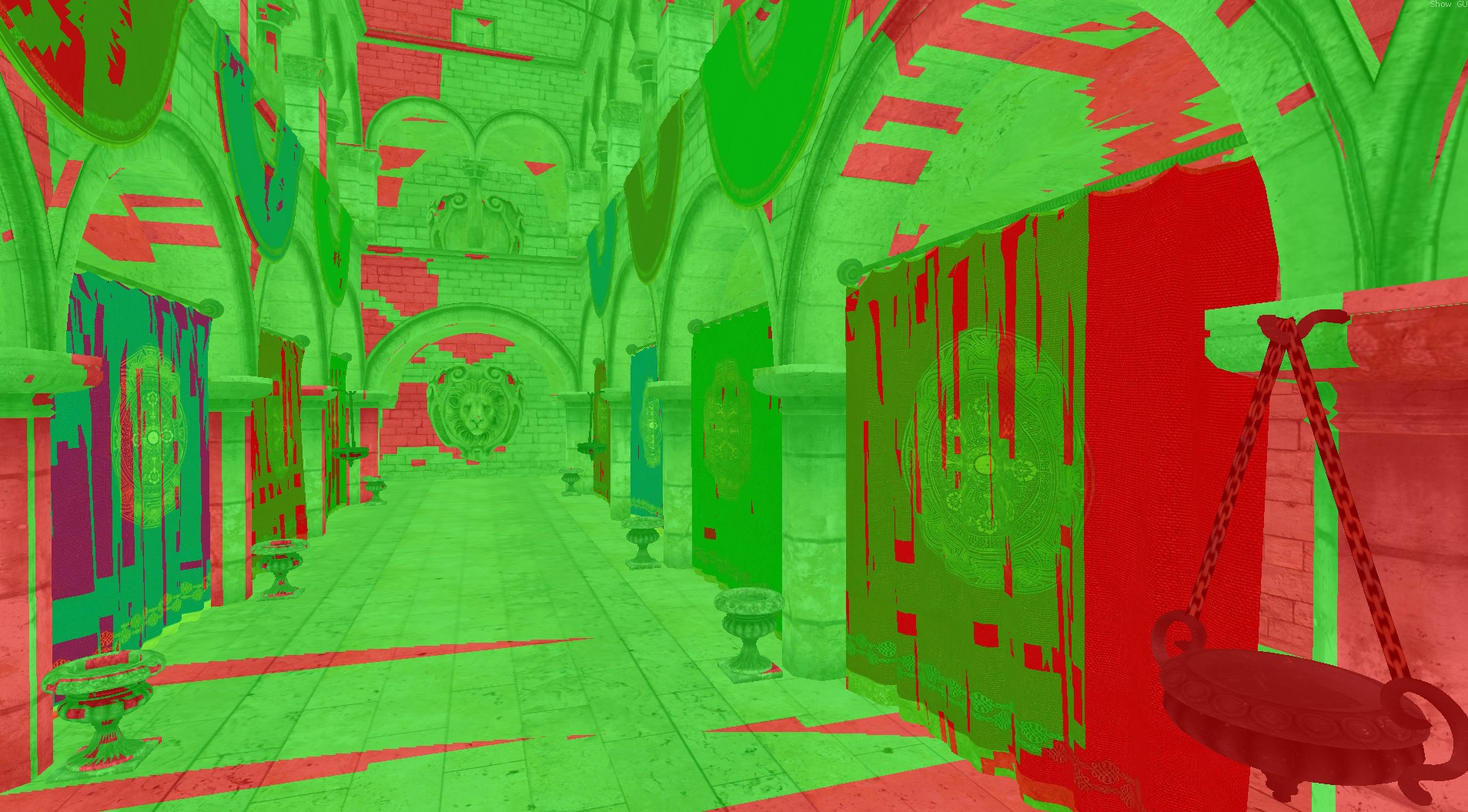}
        \caption{Green: Cached MCUs from the previous frame after a rotation by 20°. Rotating also makes MCUs from different mip map levels visible.}
        \label{fig:cache_frame}
    \end{subfigure}
    \begin{subfigure}[t]{\linewidth}
        \includegraphics[width=\linewidth]{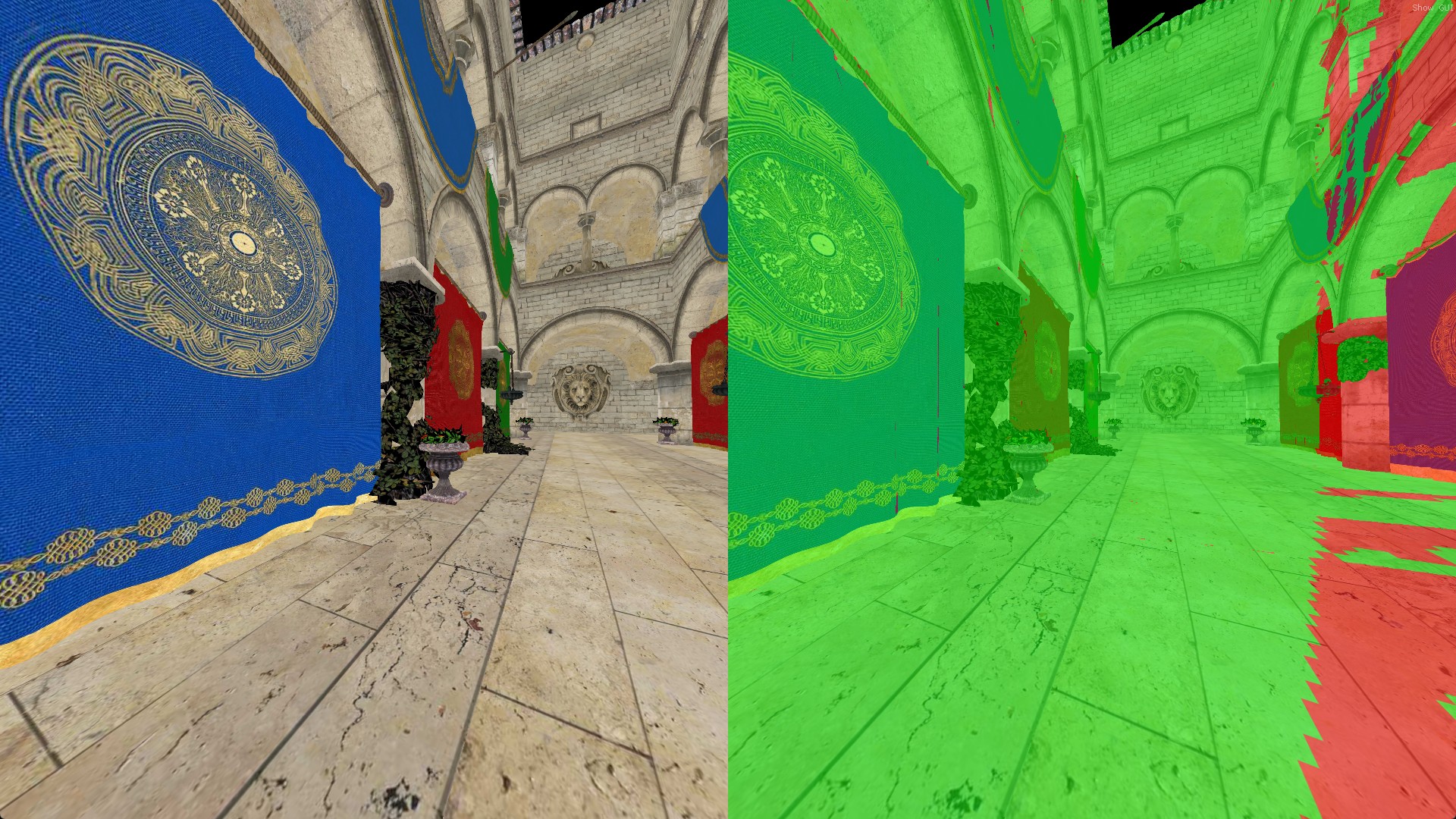}
        \caption{VR: The right eye shares most MCUs with the left. The shared frustum contains a small amount of MCUs that became visible after disocclusion. }
        \label{fig:cache_vr}
    \end{subfigure}
    \caption{Texture Block Cache: Only MCUs that were not visible in the previous frame or other VR viewpoints need decoding (red).  }
    \label{fig:cache_vis}
\end{figure}

Figure~\ref{fig:cache_vr} demonstrates that our approach scales well to virtual reality applications, as both viewpoints share the majority of MCUs. During rapid, unsophisticated head movements around Sponza, a minimum of 71\% and an average of 86\% of MCUs were shared between both eyes. In Graffiti, a minimum of 61\% and an average of 90\% of MCUs were shared. 

Applying the methodology of Section~\ref{sec:caching} to VR -- rotate by 6 degrees each frame, 60 times for a full rotation, with 100 repetitions -- we obtain a max-of-medians value of 0.65ms for the JPEG rendering pipeline in Sponza Q70, compared to the 0.20ms of the non-VR benchmark. This increase is due to the 6.5 times higher amount of pixels that are processed by the mark and resolve kernels in VR, as well as the higher number of MCUs that need decoding. Due to the larger resolution, more detailed mip map levels are needed, i.e., a larger amount of texels and corresponding MCUs are decoded. 

\subsubsection{A Note Regarding Cuda-OpenGL Interop}

We found that the first kernel launch after switching from an OpenGL context to a CUDA is unpredictably slow, adding  0.03 to 0.3 ms to the frame. Since the \emph{mark} kernel is launched first, this delay was attributed to it. To avoid this, we added a dummy kernel that is launched first and does nothing but still takes up to 0.3ms per frame. This interop cost is excluded from the JPEG pipeline timings but is inevitably reflected in the FPS reported in the teaser. 

\subsection{Memory Overhead}

A calculation of the maximum memory overhead is provided in Section~\ref{sec:overhead} (0.21 bits per texel). In practice, however, the actual overhead is lower because we counted the bits for DCs in our adaptation (12 bit) without subtracting the bits used in the original JPEG (variable). By subtracting the bits used in the original JPEG from the fixed 12-bit allocation, we obtain the following effective overheads: Sponza Q50: \textbf{0.177 bpp}; Sponza Q90: \textbf{0.154 bpp}; Graffiti Q80: \textbf{0.161 bpp}.




\section{Discussion and Future Work}

An unexpected finding of the quality evaluation (see Appendix~\ref{sec:additional_quality_results}) is that despite JPEG’s reputation as a highly efficient compression algorithm, it only modestly outperforms ASTC on tileable textures commonly used in games, according to several metrics. For photographs and non-tileable textures derived from photogrammetric mesh reconstruction, however, the quality gap is much larger and aligns with our expectations.

Recent advances like JPEG XL and AVIF demonstrate a distinctively superior trade-off between quality and size, thus we consider this JPEG-based work the first step towards adapting these latest state-of-the-art algorithms for real-time rendering. JPEG XL~\cite{sneyers2025jpegxlimagecoding}, in particular, extends many concepts of JPEG and appears well-suited for this purpose. Its advances include a more appropriate color space, predictive coding of coefficient counts, variable block sizes, and several additional features that enhance efficiency. However, some of these techniques, such as predictive coding of coefficient counts, may hinder random access because they depend on data from multiple blocks. In such cases, real-time adaptations may need to revert to alternative options, like JPEG’s end-of-block signaling in the case of coefficients.

As an alternative to mip mapping, future work could explore reconstructing lower-resolution versions of a JPEG by decoding only a subset of the AC coefficients, or exclusively the DC coefficient. This approach would allow significant reductions in computational effort at reduced resolutions without the need for explicitly stored mip levels.

In addition, due to JPEG's limitations, we currently only deal with three-channel diffuse textures without alpha. In a follow-up work, we are interested in looking at more modern standards such as JPEG XL~\cite{48554}, which promises far better compression, supports near unlimited channels and can deal with transparency. 

Virtual Texture Mapping~\cite{chajdas2018virtual} could benefit from higher compression rates, enabling faster streaming from disk to GPU. Due to minimum page sizes for disk access, latency will not be lower but larger blocks of pixels could be fetched at once. 

Neural textures may benefit from a texture block cache similar to ours, which could enable the use of more computationally intensive neural methods while maintaining real-time rendering performance. 

\section{Conclusion}


In this paper we have shown that variable-rate image compression formats such as JPEG are a viable choice for efficient real-time rendering, given a deferred rendering pipeline, a compact indexing table, mip mapping, and the use of a cache. The deferred pipeline allows us to identify the visible JPEG blocks that need decoding; the indexing table enables random access to the compressed blocks; and mip maps and cache significantly reduce the number of blocks we need to decode each frame. Although auxiliary data structures such as the indexing table add an overhead of about 0.17 bit per pixel, the resulting size is still well below the ASTC and BC formats.

At this stage, the reductions in file size of JPEG (+required overhead for random access) may not yet justify the additional computational and implementation effort in practice, but we are confident that with further development towards GPU-friendly adaptations of more recent advances like JPEG XL and AVIF, variable-rate textures will transform from a merely viable to a highly attractive choice for real-time rendering. 

Source code and data sets are available at: \url{https://github.com/elias1518693/jpeg\_textures}

\section{Acknowledgements}

The authors wish to thank following data set providers: Crytek Sponza 3D Model by Frank Meinl at Crytek, based on an older model by Marko Dabrović; Photograph of the Kiyomizu-Dera temple by Thomas Rausch; Textures provided by https://ambientcg.com/ and https://polyhaven.com/.

This research has been funded by WWTF project \emph{ICT22-055 - Instant Visualization and Interaction for Large Point Clouds}. 

\printbibliography                

@inproceedings{ASTC,
author = {Nystad, J. and Lassen, A. and Pomianowski, A. and Ellis, S. and Olson, T.},
title = {Adaptive scalable texture compression},
year = {2012},
isbn = {9783905674415},
publisher = {Eurographics Association},
address = {Goslar, DEU},
booktitle = {Proceedings of the Fourth ACM SIGGRAPH / Eurographics Conference on High-Performance Graphics},
pages = {105–114},
numpages = {10},
location = {Paris, France},
series = {EGGH-HPG'12}
}

@inproceedings {10.2312:hpg.20211284,
    booktitle = {High-Performance Graphics - Symposium Papers},
    editor = {Binder, Nikolaus and Ritschel, Tobias},
    title = {{Compression and Rendering of Textured Point Clouds via Sparse Coding}},
    author = {Schuster, Kersten and Trettner, Philip and Schmitz, Patric and Schakib, Julian and Kobbelt, Leif},
    year = {2021},
    publisher = {The Eurographics Association},
    ISSN = {2079-8687},
    ISBN = {978-3-03868-156-4},
    DOI = {10.2312/hpg.20211284}
}

@inproceedings{ntc2023,
    author = {Vaidyanathan, Karthik and Salvi, Marco and Wronski, Bartlomiej and Akenine-Möller, Tomas and Ebelin, Pontus and Lefohn, Aaron},
    title = "{Random-Access Neural Compression of Material Textures}",
    year = {2023},
    booktitle = {Proceedings of SIGGRAPH},
}

@article{10.1111:j.1467-8659.2011.01989.x,
journal = {Computer Graphics Forum},
title = {{Variable Bit Rate GPU Texture Decompression}},
author = {Olano, Marc and Baker, Dan and Griffin, Wesley and Barczak, Joshua},
year = {2011},
publisher = {The Eurographics Association and Blackwell Publishing Ltd.},
ISSN = {1467-8659},
DOI = {10.1111/j.1467-8659.2011.01989.x}
}

@article{radziszewski2008optimization,
  title={Optimization of frequency filtering in random access JPEG library},
  author={Radziszewski, Michal and Alda, Witold},
  journal={Computer Science},
  volume={9},
  pages={109--120},
  year={2008},
  publisher={Akademia G{\'o}rniczo-Hutnicza im. Stanis{\l}awa Staszica w Krakowie. Wydawnictwo AGH}
}

@patent{Iourcha1999S3TC,
  author = {Iourcha, Konstantine and Nayak, Krishna S. and Hong, Zhou},
  title = {System and method for fixed-rate block-based image compression with inferred pixel values},
  number = {US5956431},
  type = {Patent},
  year = {1999},
  month = {Sep},
  day = {21},
  holder = {S3 Incorporated},
  url = {https://patents.google.com/patent/US5956431}
}

@inproceedings{10.1145/3610548.3618150,
    author = {Luo, Yuzhe and Jin, Xiaogang and Pan, Zherong and Wu, Kui and Kou, Qilong and Yang, Xiajun and Gao, Xifeng},
    title = {Texture Atlas Compression Based on Repeated Content Removal},
    year = {2023},
    isbn = {9798400703157},
    publisher = {Association for Computing Machinery},
    address = {New York, NY, USA},
    doi = {10.1145/3610548.3618150},
    booktitle = {SIGGRAPH Asia 2023 Conference Papers},
    articleno = {52},
    numpages = {11},
    location = {Sydney, NSW, Australia},
    series = {SA '23}
}

@ARTICLE{125072,
  author={Wallace, G.K.},
  journal={IEEE Transactions on Consumer Electronics}, 
  title={The JPEG still picture compression standard}, 
  year={1992},
  volume={38},
  number={1},
  pages={xviii-xxxiv},
  doi={10.1109/30.125072}}

@article{https://doi.org/10.2312/mam.20241178,
    doi = {10.2312/MAM.20241178},
    author = {Fujieda, Shin and Harada, Takahiro},
    title = {Neural Texture Block Compression},
    publisher = {The Eurographics Association},
    year = {2024},
    copyright = {Creative Commons Attribution 4.0 International}
}

@misc{farhadzadeh2024neuralgraphicstexturecompression,
      title={Neural Graphics Texture Compression Supporting Random Access}, 
      author={Farzad Farhadzadeh and Qiqi Hou and Hoang Le and Amir Said and Randall Rauwendaal and Alex Bourd and Fatih Porikli},
      year={2024},
      eprint={2407.00021},
      archivePrefix={arXiv},
      primaryClass={cs.CV},
      url={https://arxiv.org/abs/2407.00021}, 
}

@ARTICLE{4051119,
  author={Huffman, David A.},
  journal={Proceedings of the IRE}, 
  title={A Method for the Construction of Minimum-Redundancy Codes}, 
  year={1952},
  volume={40},
  number={9},
  pages={1098-1101},
  doi={10.1109/JRPROC.1952.273898}}

@article{Chen2002AJT,
  title={A JPEG-like texture compression with adaptive quantization for 3D graphics application},
  author={C.-H. Chen and C.-Y. Lee},
  journal={The Visual Computer},
  year={2002},
  volume={18},
  pages={29-40},
  url={https://api.semanticscholar.org/CorpusID:8072089}
}

@article{10.1007/s00371-011-0621-8,
    author = {Hollemeersch, Charles-Frederik and Pieters, Bart and Lambert, Peter and Van de Walle, Rik},
    title = {A new approach to combine texture compression and filtering},
    year = {2012},
    issue_date = {April 2012},
    publisher = {Springer-Verlag},
    address = {Berlin, Heidelberg},
    volume = {28},
    number = {4},
    issn = {0178-2789},
    doi = {10.1007/s00371-011-0621-8},
    journal = {Vis. Comput.},
    month = apr,
    pages = {371–385},
    numpages = {15}
}

@inproceedings{48554,title	= {
    JPEG XL next-generation image compression architecture and coding tools},
    author	= {Jyrki Alakuijala and Ruud van Asseldonk and Sami Boukortt and Martin Bruse and Iulia-Maria Comsa and Moritz Firsching and Thomas Fischbacher and Sebastian Gomez and Evgenii Kliuchnikov and Robert Obryk and Krzysztof Potempa and Alexander Rhatushnyak and Jon Sneyers and Zoltan Szabadka and Lode Vandevenne and Luca Versari and Jan Wassenberg},
    year	= {2019},
    URL	= {https://www.spiedigitallibrary.org/conference-proceedings-of-spie/11137/111370K/JPEG-XL-next-generation-image-compression-architecture-and-coding-tools/10.1117/12.2529237.full}
}

@inproceedings{10.1145/1071866.1071877,
    author = {Str\"{o}m, Jacob and Akenine-M\"{o}ller, Tomas},
    title = {iPACKMAN: high-quality, low-complexity texture compression for mobile phones},
    year = {2005},
    isbn = {1595930868},
    publisher = {Association for Computing Machinery},
    address = {New York, NY, USA},
    doi = {10.1145/1071866.1071877},
    booktitle = {Proceedings of the ACM SIGGRAPH/EUROGRAPHICS Conference on Graphics Hardware},
    pages = {63–70},
    numpages = {8},
    location = {Los Angeles, California},
    series = {HWWS '05}
}

@inproceedings{10.1145/357744.357757,
    author = {Christopoulos, C. A. and Ebrahimi, T. and Skodras, A. N.},
    title = {JPEG2000: the new still picture compression standard},
    year = {2000},
    isbn = {1581133111},
    publisher = {Association for Computing Machinery},
    address = {New York, NY, USA},
    doi = {10.1145/357744.357757},
    booktitle = {Proceedings of the 2000 ACM Workshops on Multimedia},
    pages = {45–49},
    numpages = {5},
    location = {Los Angeles, California, USA},
    series = {MULTIMEDIA '00}
}

@article{andersson2020flip,
  title={FLIP: A Difference Evaluator for Alternating Images.},
  author={Andersson, Pontus and Nilsson, Jim and Akenine-M{\"o}ller, Tomas and Oskarsson, Magnus and {\AA}str{\"o}m, Kalle and Fairchild, Mark D},
  journal={Proc. ACM Comput. Graph. Interact. Tech.},
  volume={3},
  number={2},
  pages={15--1},
  year={2020}
}

@online{McGuire2017Data,
  title = {Computer Graphics Archive},
  author = {Morgan McGuire},
  year = {2017},
  month = {July},
  url = {https://casual-effects.com/data}
}

@article{fichet2025compression,
  title={Compression of Spectral Images Using Spectral JPEG XL},
  author={Fichet, Alban and Peters, Christoph},
  journal={Journal of Computer Graphics Techniques Vol},
  volume={14},
  number={1},
  year={2025}
}

@ARTICLE{SSIM,
    author={Zhou Wang and Bovik, A.C. and Sheikh, H.R. and Simoncelli, E.P.},
    journal={IEEE Transactions on Image Processing}, 
    title={Image quality assessment: from error visibility to structural similarity}, 
    year={2004},
    volume={13},
    number={4},
    pages={600-612},
    doi={10.1109/TIP.2003.819861}
}

@inproceedings{LPIPS,
  title={The Unreasonable Effectiveness of Deep Features as a Perceptual Metric},
  author={Zhang, Richard and Isola, Phillip and Efros, Alexei A and Shechtman, Eli and Wang, Oliver},
  booktitle={CVPR},
  year={2018}
}

@misc{sneyers2025jpegxlimagecoding,
    title={The JPEG XL Image Coding System: History, Features, Coding Tools, Design Rationale, and Future}, 
    author={Jon Sneyers and Jyrki Alakuijala and Luca Versari and Zoltán Szabadka and Sami Boukortt and Amnon Cohen-Tidhar and Moritz Firsching and Evgenii Kliuchnikov and Tal Lev-Ami and Eric Portis and Thomas Richter and Osamu Watanabe},
    year={2025},
    eprint={2506.05987},
    archivePrefix={arXiv},
    primaryClass={cs.MM},
}

@misc{laurent2025hardwareacceleratedneuralblock,
      title={Hardware Accelerated Neural Block Texture Compression with Cooperative Vectors}, 
      author={Belcour Laurent and Benyoub Anis},
      year={2025},
      eprint={2506.06040},
      archivePrefix={arXiv},
      primaryClass={cs.GR},
}

@misc{weinreich2024realtimeneuralmaterialsusing,
      title={Real-Time Neural Materials using Block-Compressed Features}, 
      author={Clément Weinreich and Louis de Oliveira and Antoine Houdard and Georges Nader},
      year={2024},
      eprint={2311.16121},
      archivePrefix={arXiv},
      primaryClass={cs.CV},
      url={https://arxiv.org/abs/2311.16121}, 
}

@inproceedings{zhang2025image,
  title={Image-gs: Content-adaptive image representation via 2d gaussians},
  author={Zhang, Yunxiang and Li, Bingxuan and Kuznetsov, Alexandr and Jindal, Akshay and Diolatzis, Stavros and Chen, Kenneth and Sochenov, Anton and Kaplanyan, Anton and Sun, Qi},
  booktitle={Proceedings of the Special Interest Group on Computer Graphics and Interactive Techniques Conference Conference Papers},
  pages={1--11},
  year={2025}
}

@Article{kerbl3Dgaussians,
      author       = {Kerbl, Bernhard and Kopanas, Georgios and Leimk{\"u}hler, Thomas and Drettakis, George},
      title        = {3D Gaussian Splatting for Real-Time Radiance Field Rendering},
      journal      = {ACM Transactions on Graphics},
      number       = {4},
      volume       = {42},
      month        = {July},
      year         = {2023},
      url          = {https://repo-sam.inria.fr/fungraph/3d-gaussian-splatting/}
}

@misc{zhang2024gaussianimage1000fpsimage,
      title={GaussianImage: 1000 FPS Image Representation and Compression by 2D Gaussian Splatting}, 
      author={Xinjie Zhang and Xingtong Ge and Tongda Xu and Dailan He and Yan Wang and Hongwei Qin and Guo Lu and Jing Geng and Jun Zhang},
      year={2024},
      eprint={2403.08551},
      archivePrefix={arXiv},
      primaryClass={eess.IV},
      url={https://arxiv.org/abs/2403.08551}, 
}

@incollection{chajdas2018virtual,
    title={Virtual texture mapping 101},
    author={Chajdas, Matth{\"a}us G and Eisenacher, Christian and Stamminger, Marc and Lefebvre, Sylvain},
    booktitle={GPU Pro 360 Guide to Rendering},
    pages={69--79},
    year={2018},
    publisher={AK Peters/CRC Press}
}


\onecolumn
\appendix
\section{Additional Quality Evaluation Results}
\label{sec:additional_quality_results}


\setlength{\tabcolsep}{1pt}
\renewcommand{\arraystretch}{0.5}
\begin{table}[H]
\caption{Artifacts of various compression algorithms at a target bitrate of 0.89 bit per pixel (bpp), or the quality level that comes closest. BC1 for comparison (always 4bpp). Each cell shows quality and bpp in the top row, and FLIP and PSNR in the bottom row. Best and worst FLIP and PSNR values in a row are highlighted, with the exception of BC1 which is not considered due to its high memory usage.}
\begin{tabular*}{\textwidth}{cccccc}
  \textbf{Reference}  & \textbf{BC1} & \textbf{ASTC 12x12} & \textbf{JPEG} & \textbf{JPEG XL} & \textbf{AVIF} \\ 
  \qcellc{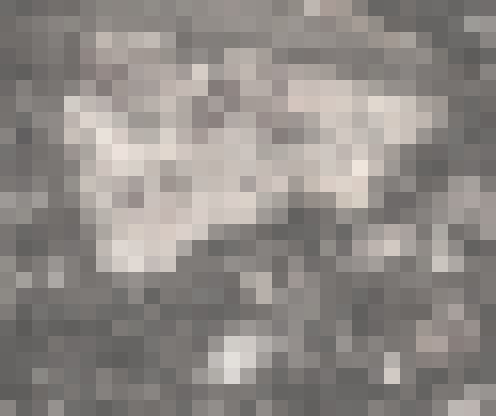} {asphalt\_04\_diff\_4k}{}{}{} & 
  \qcellc{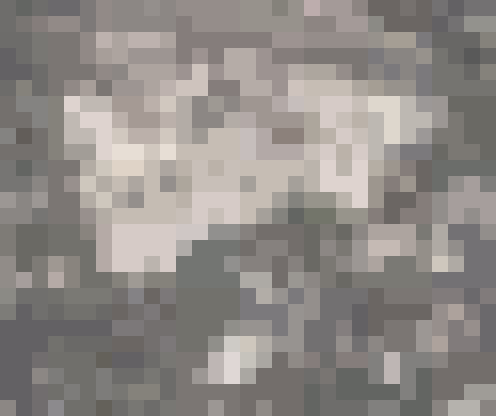}                {   }{4.00bpp}{0.048}            {34.1} & 
  \qcellc{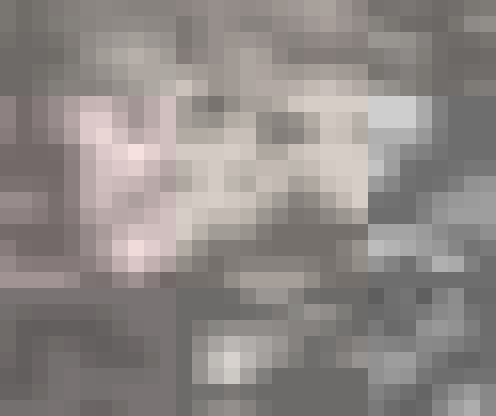}     {   }{0.89bpp}{\loser{0.108}}    {26.0} & 
  \qcellc{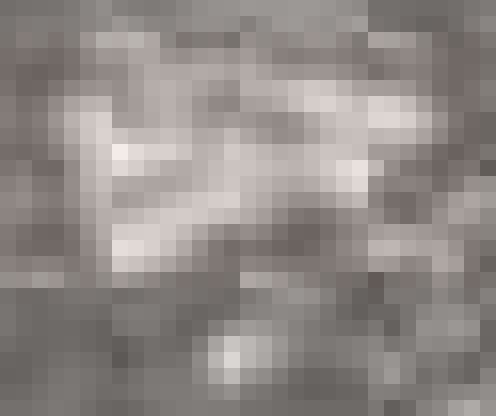}          {Q40}{0.86bpp}{0.071}            {\loser{25.8}} & 
  \qcellc{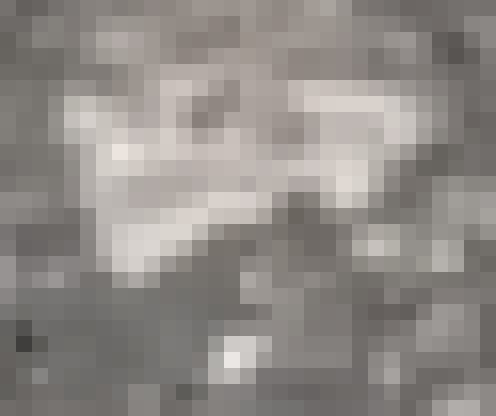}          {Q65}{0.82bpp}{\winner{0.069}}   {26.9} & 
  \qcellc{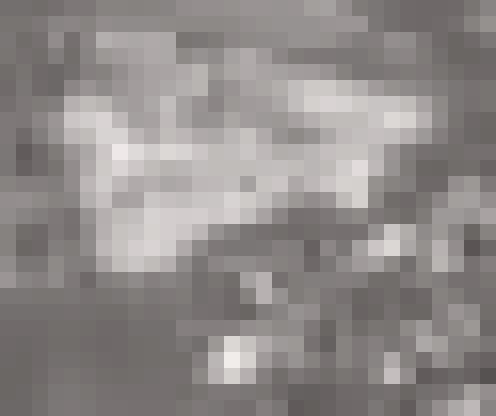}            {Q40}{0.98bpp}{0.082}            {\winner{28.4}} \\
  \qcellc{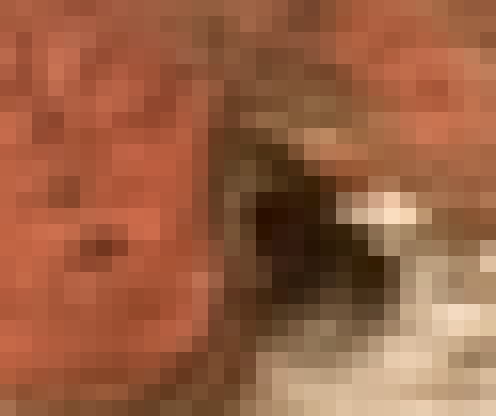}{brick\_wall\_006}{}{}{} &
  \qcellc{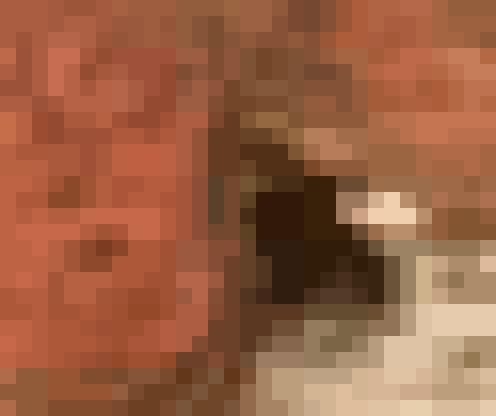}                   {   }{4.00bpp}{0.048}{35.3} &
  \qcellc{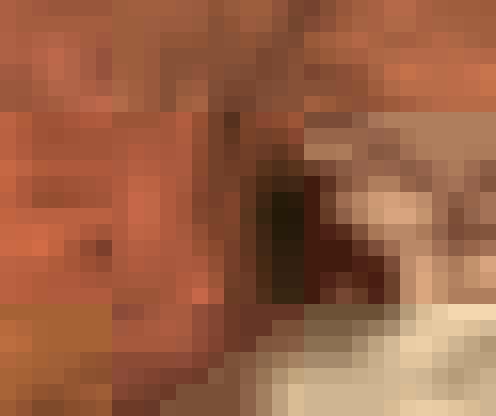}        {   }{0.89bpp}{\loser{0.104}}{\loser{28.8}} &
  \qcellc{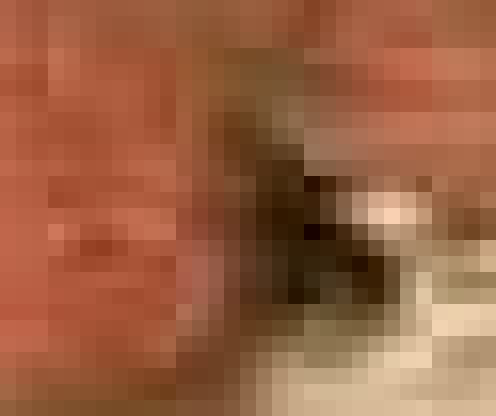}             {Q50}{0.91bpp}{0.091}{29.3} &
  \qcellc{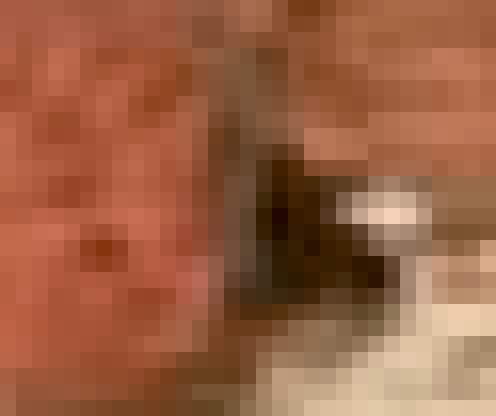}             {Q70}{0.87bpp}{0.080}{30.5} &
  \qcellc{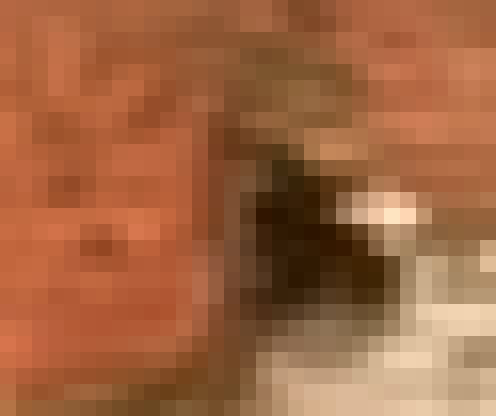}               {Q50}{0.93bpp}{\winner{0.079}}{\winner{32.1}}\\
  \qcellc{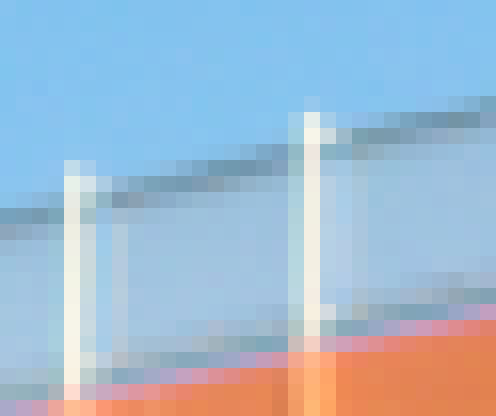} {bridge\_1024\_768}{}{}{} & 
  \qcellc{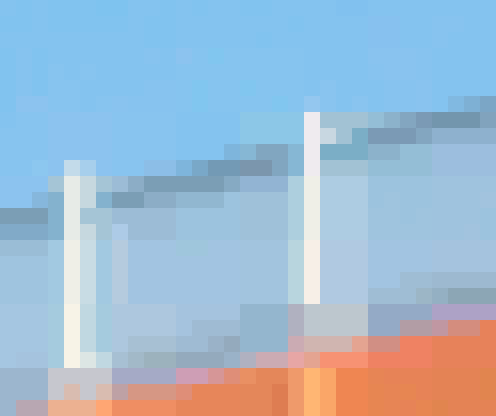}             {   }{4.00bpp}{0.044}{34.6} & 
  \qcellc{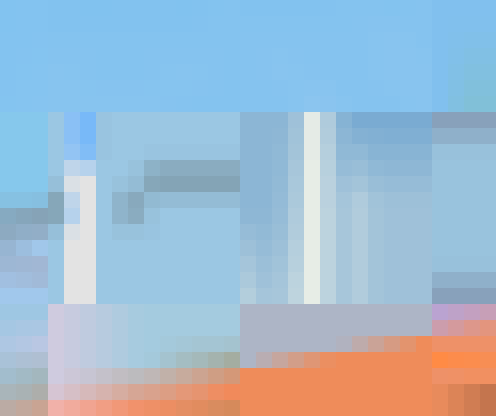}  {   }{0.89bpp}{\loser{0.076}}{\loser{29.4}} & 
  \qcellc{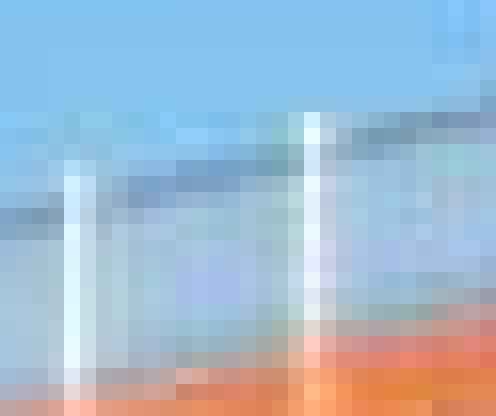}       {Q70}{0.93bpp}{0.063}{30.5} & 
  \qcellc{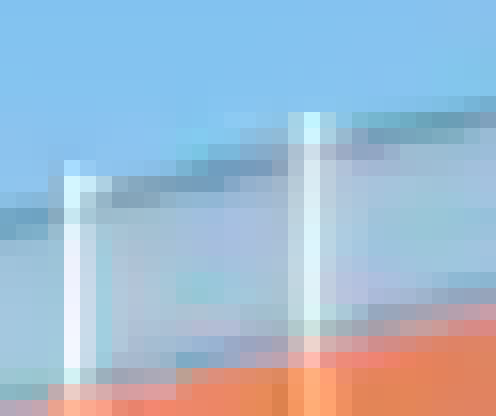}       {Q75}{0.77bpp}{\winner{0.054}}{31.2} & 
  \qcellc{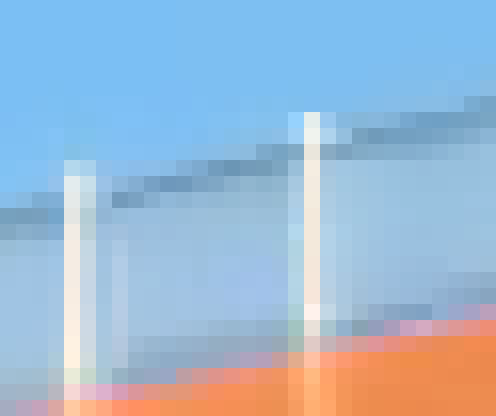}         {Q55}{0.86bpp}{0.056}{\winner{33.9}} \\
  \qcellc{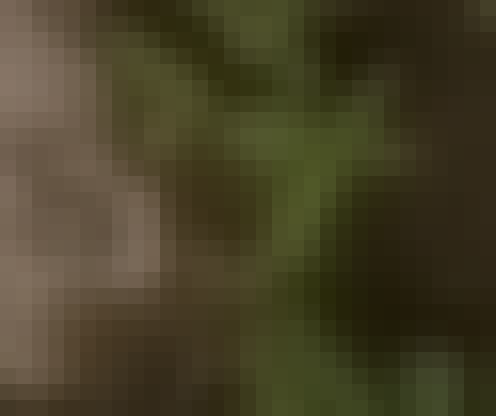}     {coast\_sand}{}{}{} &
  \qcellc{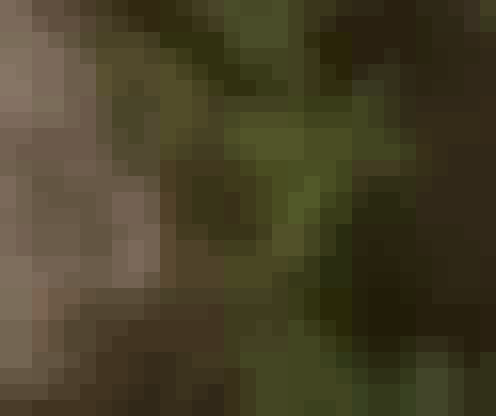}            {   }{4.00bpp}{0.028}{42.3} &
  \qcellc{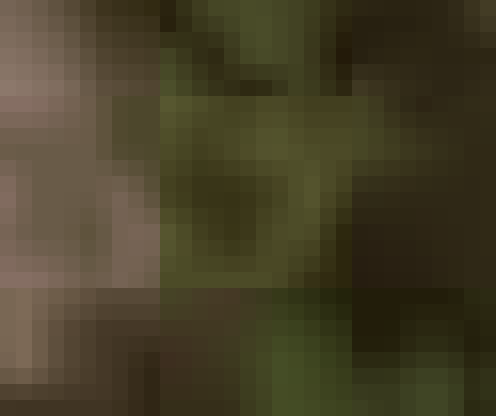} {   }{0.89bpp}{\loser{0.044}}{\loser{39.9}} &
  \qcellc{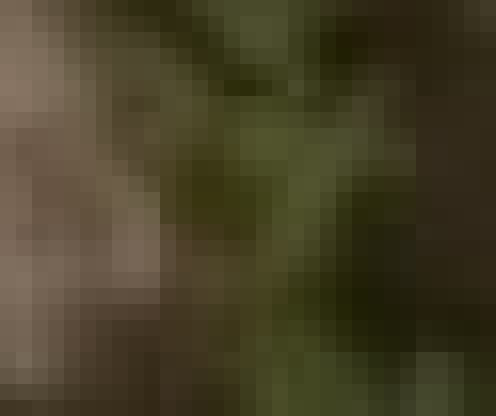}      {Q85}{0.90bpp}{0.037}{41.1} &
  \qcellc{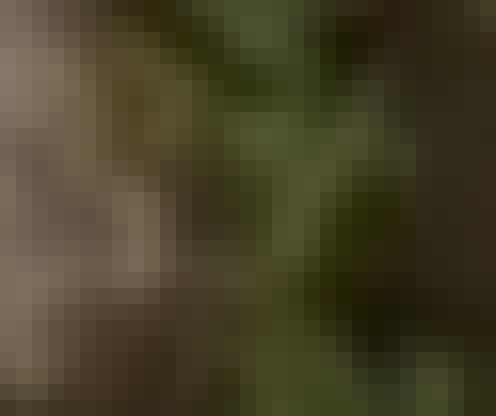}      {Q85}{0.81bpp}{\winner{0.029}}{43.3} &
  \qcellc{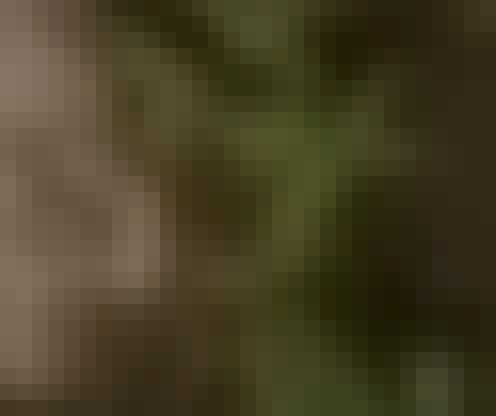}        {Q85}{0.86bpp}{\winner{0.029}}{\winner{44.2}} \\
  \qcellc{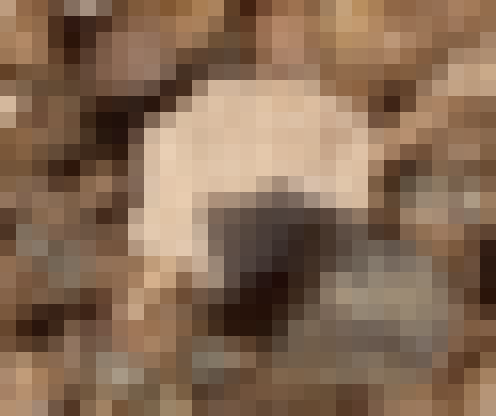}{coral\_gravel\_diff\_4k}{}{}{} &
  \qcellc{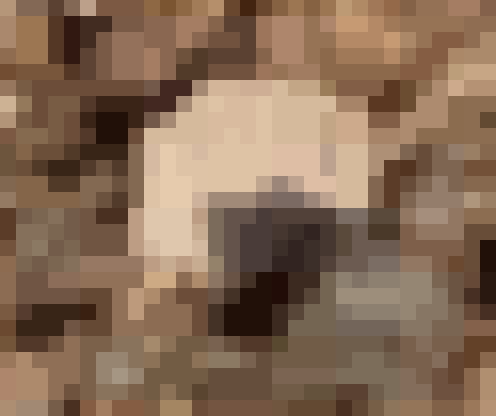}                 {   }{4.00bpp}{0.051}{33.4} &
  \qcellc{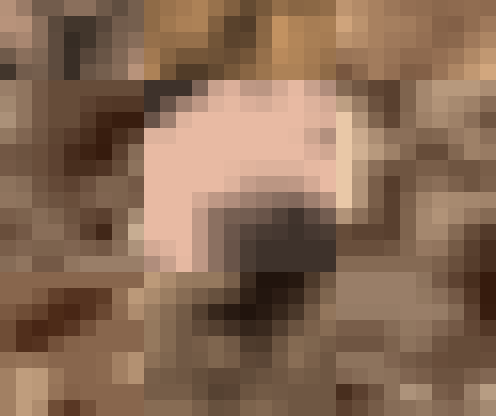}      {   }{0.89bpp}{\loser{0.113}}{\loser{27.1}} &
  \qcellc{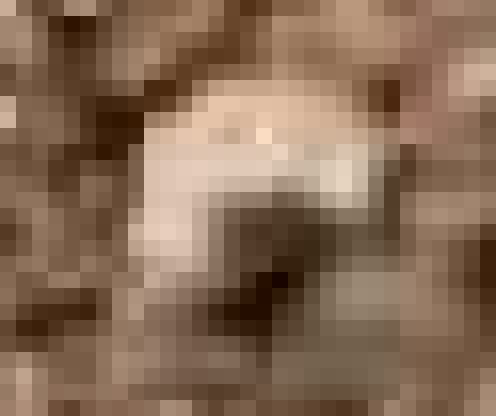}           {Q40}{0.99bpp}{0.094}{27.9} &
  \qcellc{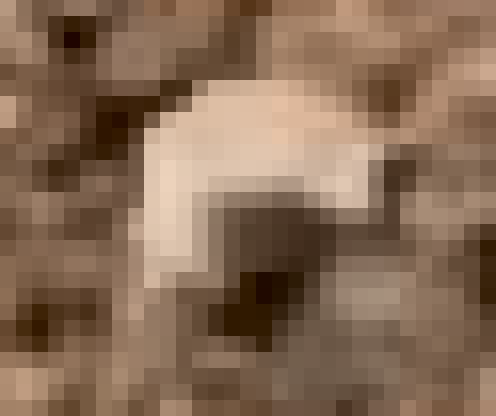}           {Q65}{0.90bpp}{\winner{0.082}}{29.3} &
  \qcellc{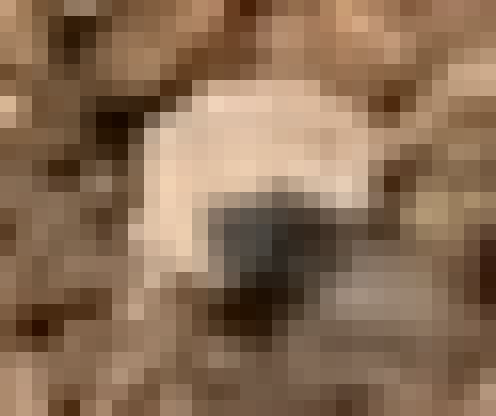}             {Q45}{0.98bpp}{0.084}{\winner{30.7}} \\
  \qcellc{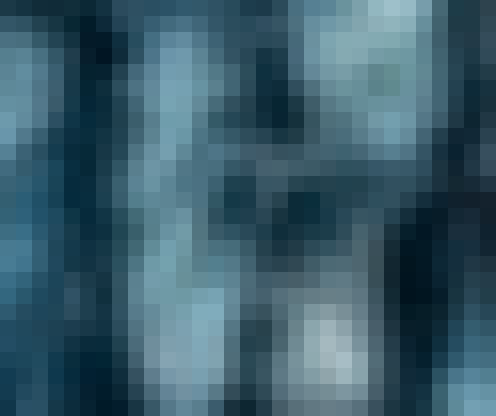}{denmin\_fabric\_02}{}{}{} &
  \qcellc{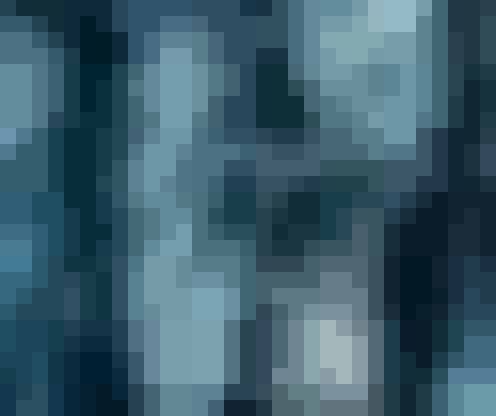}                 {   }{4.00bpp}{0.43}{33.7} &
  \qcellc{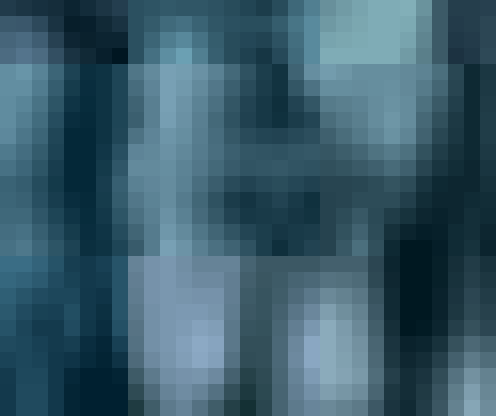}      {   }{0.89bpp}{\loser{0.089}}{\loser{29.8}} &
  \qcellc{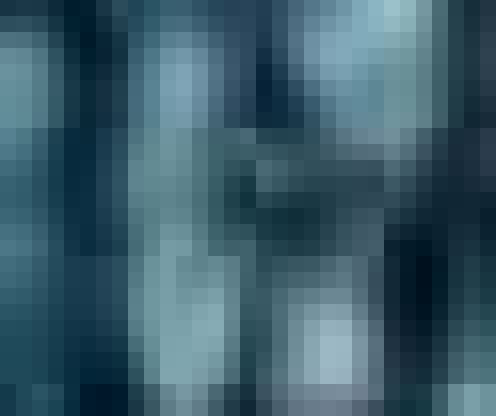}           {Q40}{0.94bpp}{0.079}{30.6} &
  \qcellc{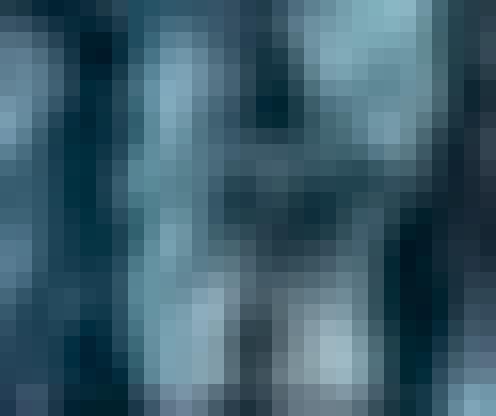}           {Q74}{0.83bpp}{0.061}{32.6} &
  \qcellc{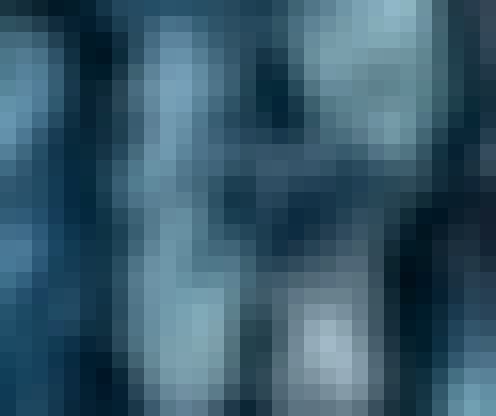}             {Q55}{0.95bpp}{\winner{0.059}}{\winner{34.7}} \\
  \qcellc{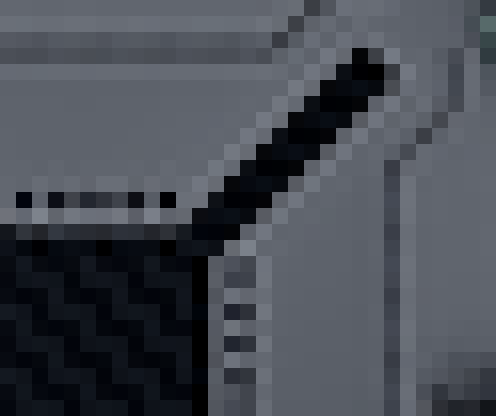}{Chip005}{}{}{} &
  \qcellc{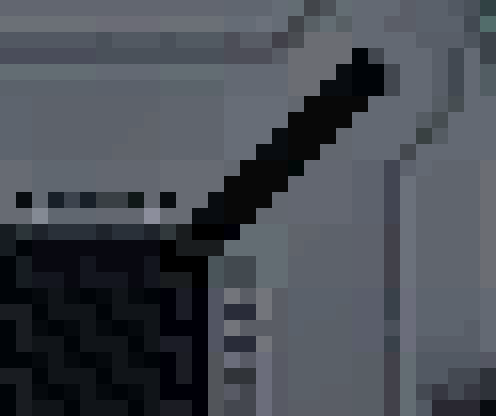}                 {   }{4.00bpp}{0.036}{34.6} &
  \qcellc{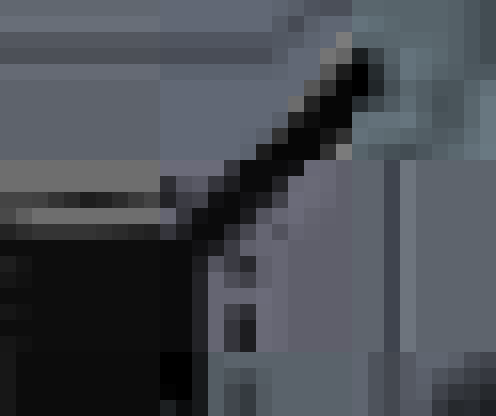}      {   }{0.89bpp}{\loser{0.076}}{\loser{27.0}} &
  \qcellc{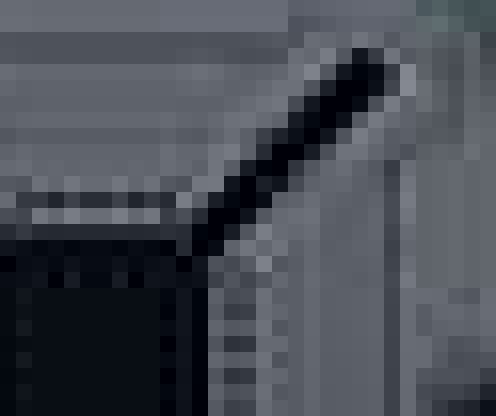}           {Q55}{0.93bpp}{0.065}{27.7} &
  \qcellc{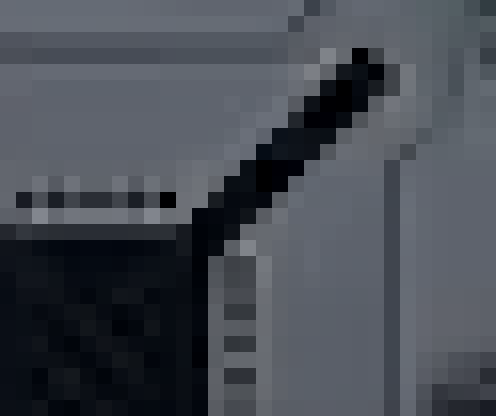}           {Q70}{0.93bpp}{0.058}{29.6} &
  \qcellc{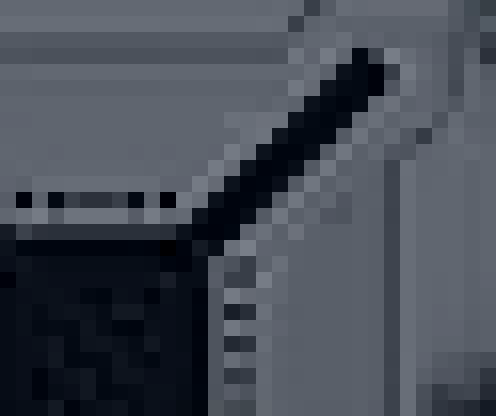}             {Q50}{0.95bpp}{\winner{0.055}}{\winner{32.4}} \\
\end{tabular*}
\label{tab:appendix_artifacts}
\end{table}

\newcommand{\IMGWIDTH}{0.249}

\begin{table}[]
\caption{Quality metric plots for the test data sets in Table~\ref{tab:appendix_artifacts}. }
\begin{tabular*}{\textwidth}{cccc}
    PSNR↑ & FLIP↓ & LPIPS↓ & SSIM↑ \\ 
    \plotrow{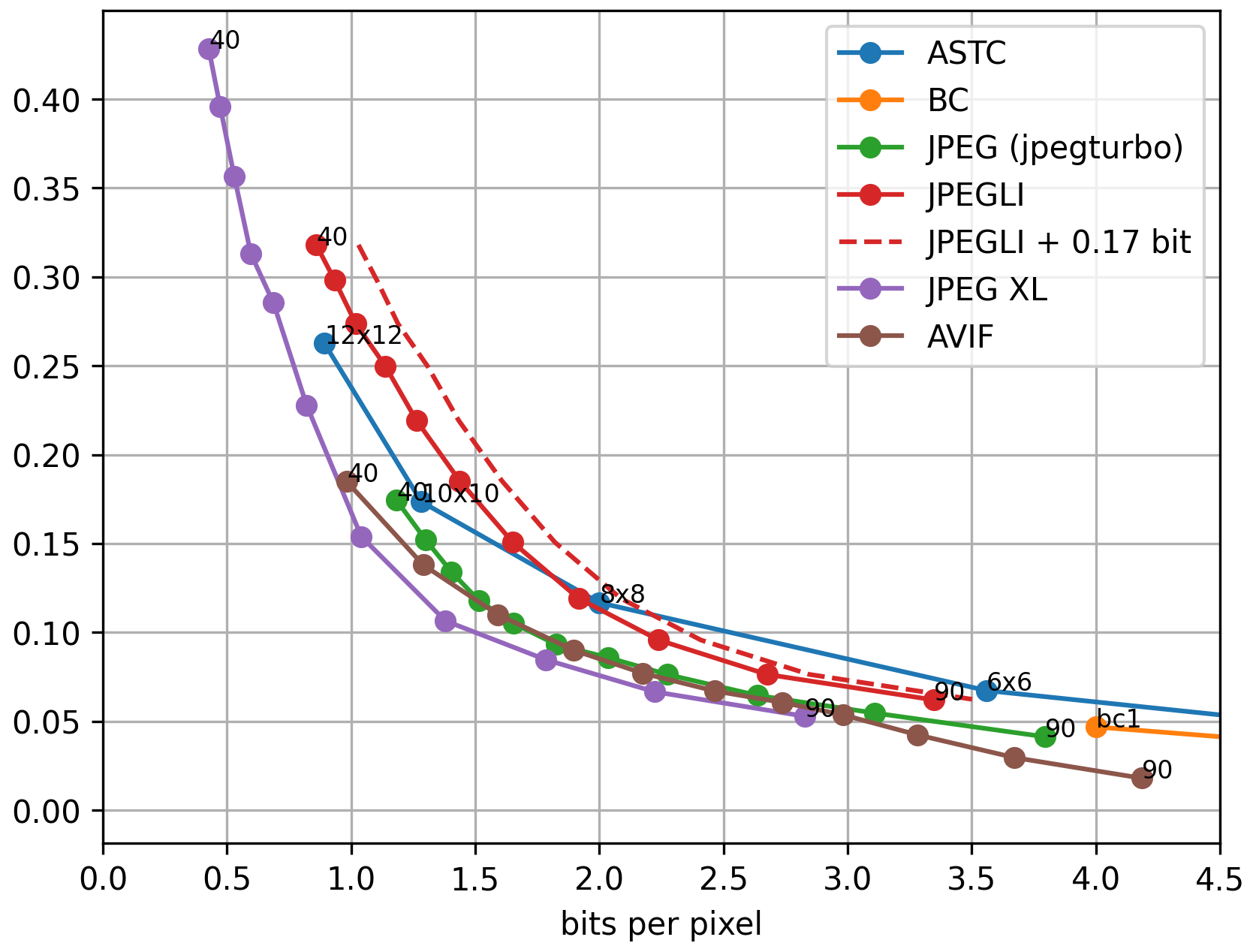}{images/qualitative/asphalt_04_diff_4k.png/asphalt_04_diff_4k.jpg}{34px}{0.3pt}
    \plotrow{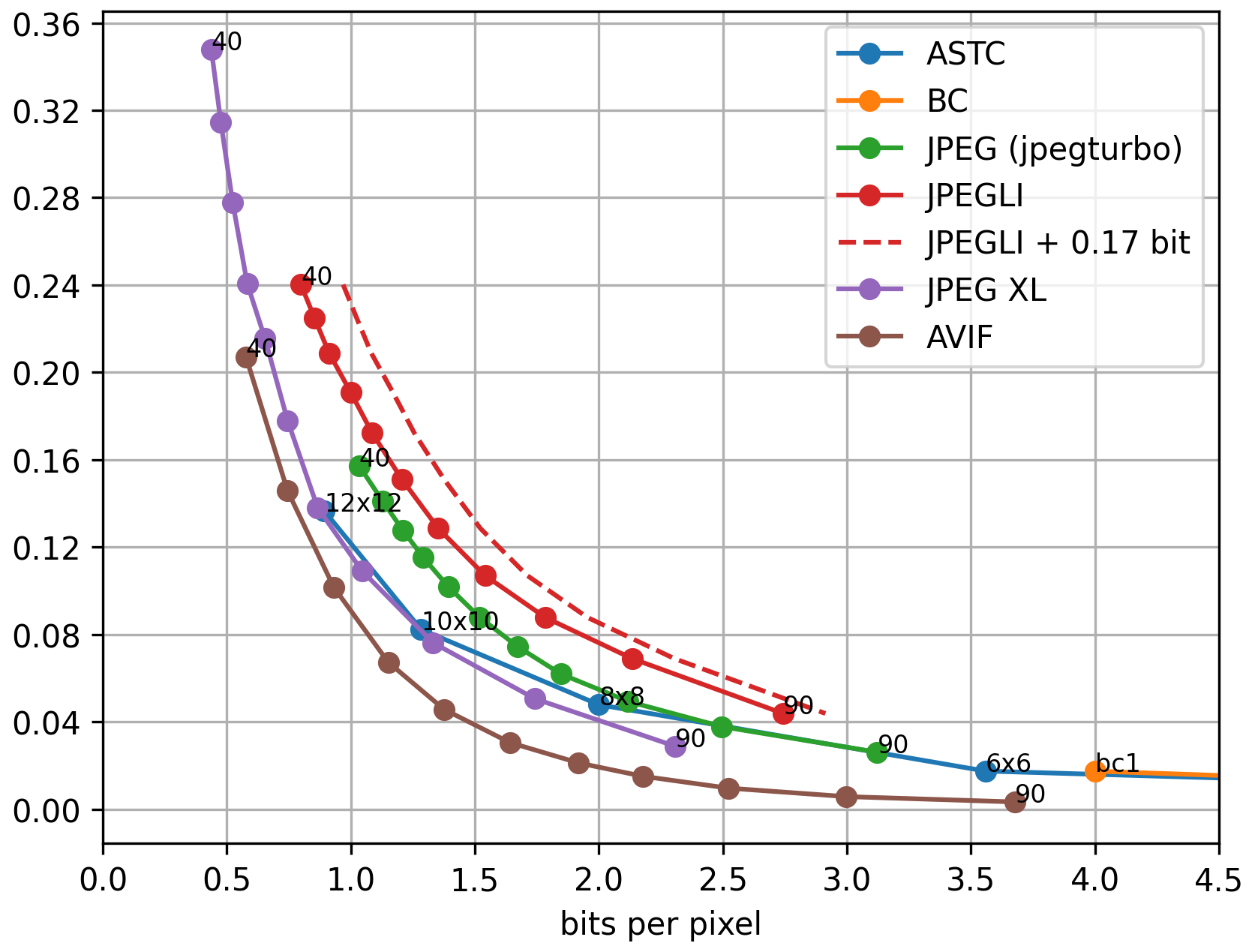}{images/qualitative/brick_wall_006_diff_4k.png/brick_wall_006_diff_4k.jpg}{34px}{0.3pt}
    \plotrow{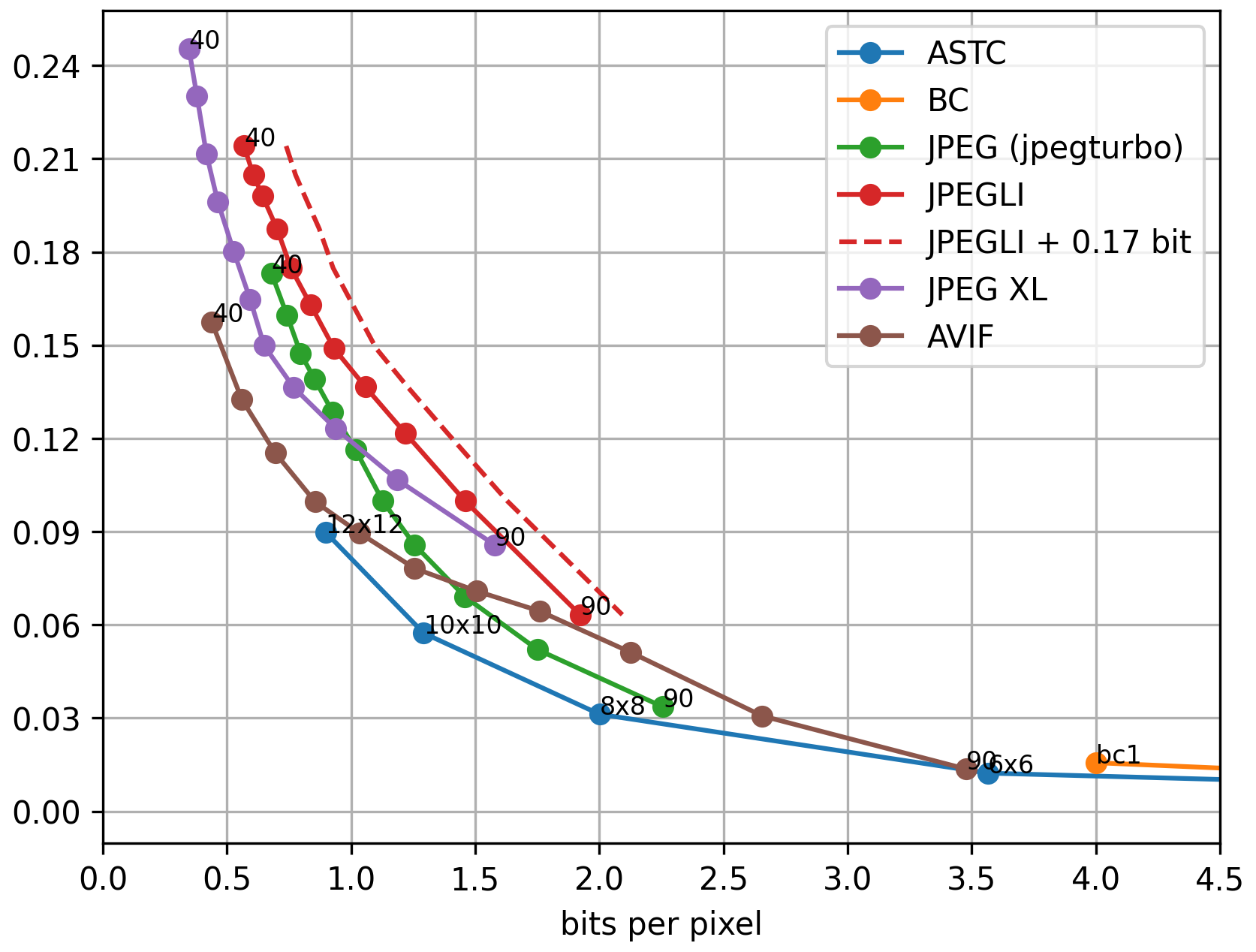}{images/qualitative/bridge_1024_768.png/bridge_1024_768.jpg}{34px}{0.3pt}
    \plotrow{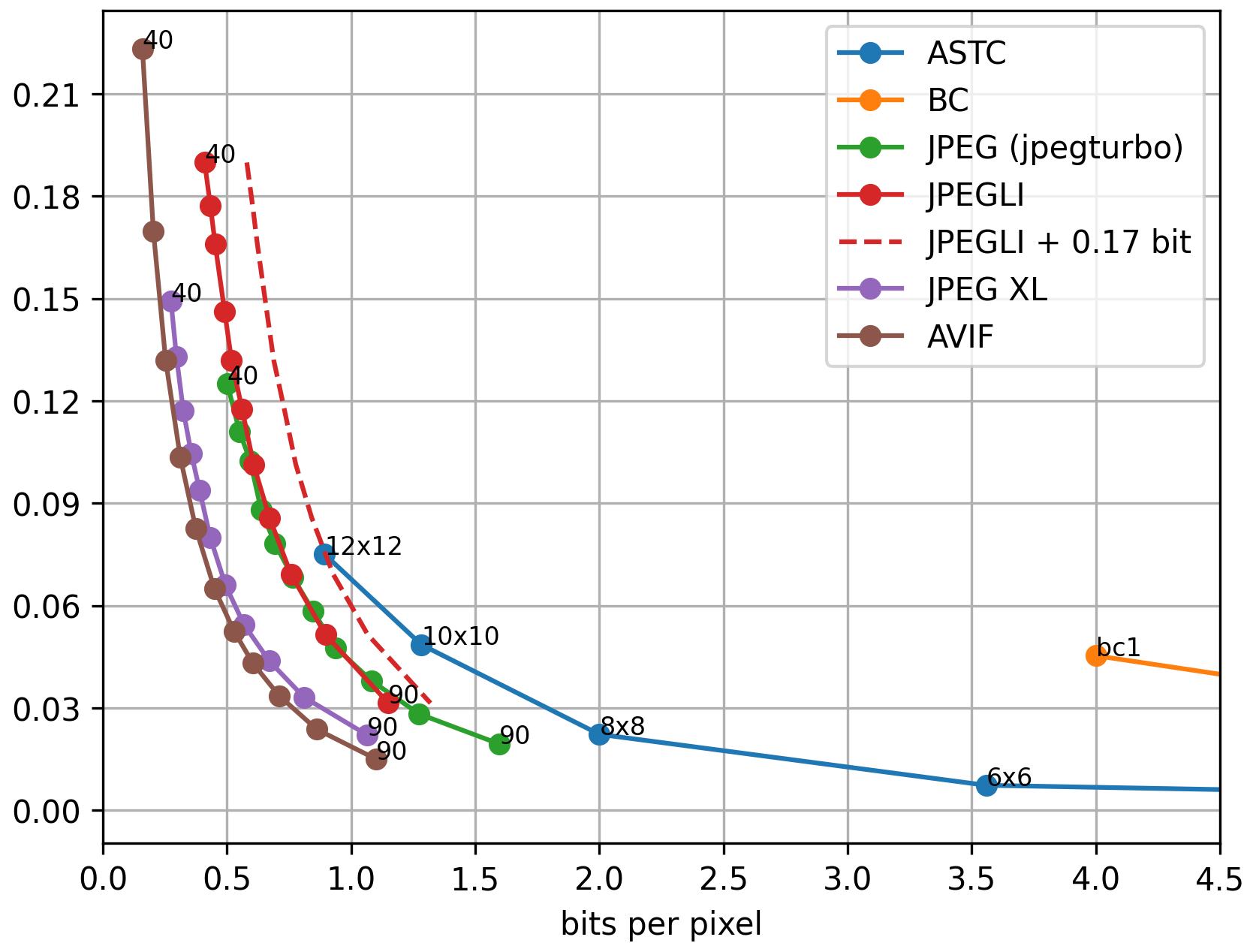}{images/qualitative/coast_sand.png/coast_sand.jpg}{34px}{0.3pt}
    \plotrow{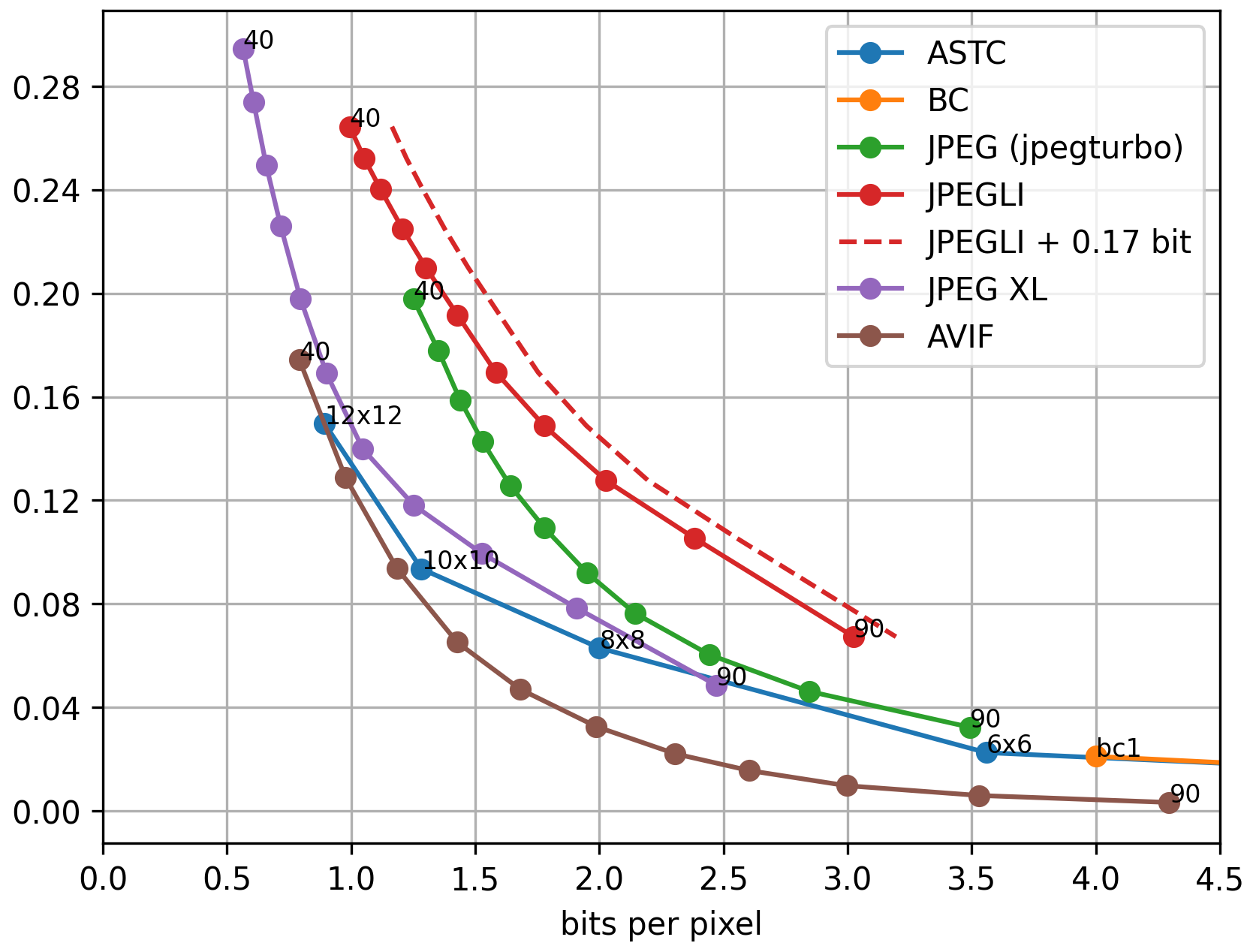}{images/qualitative/coral_gravel_diff_4k.png/coral_gravel_diff_4k.jpg}{34px}{0.3pt}
    \plotrow{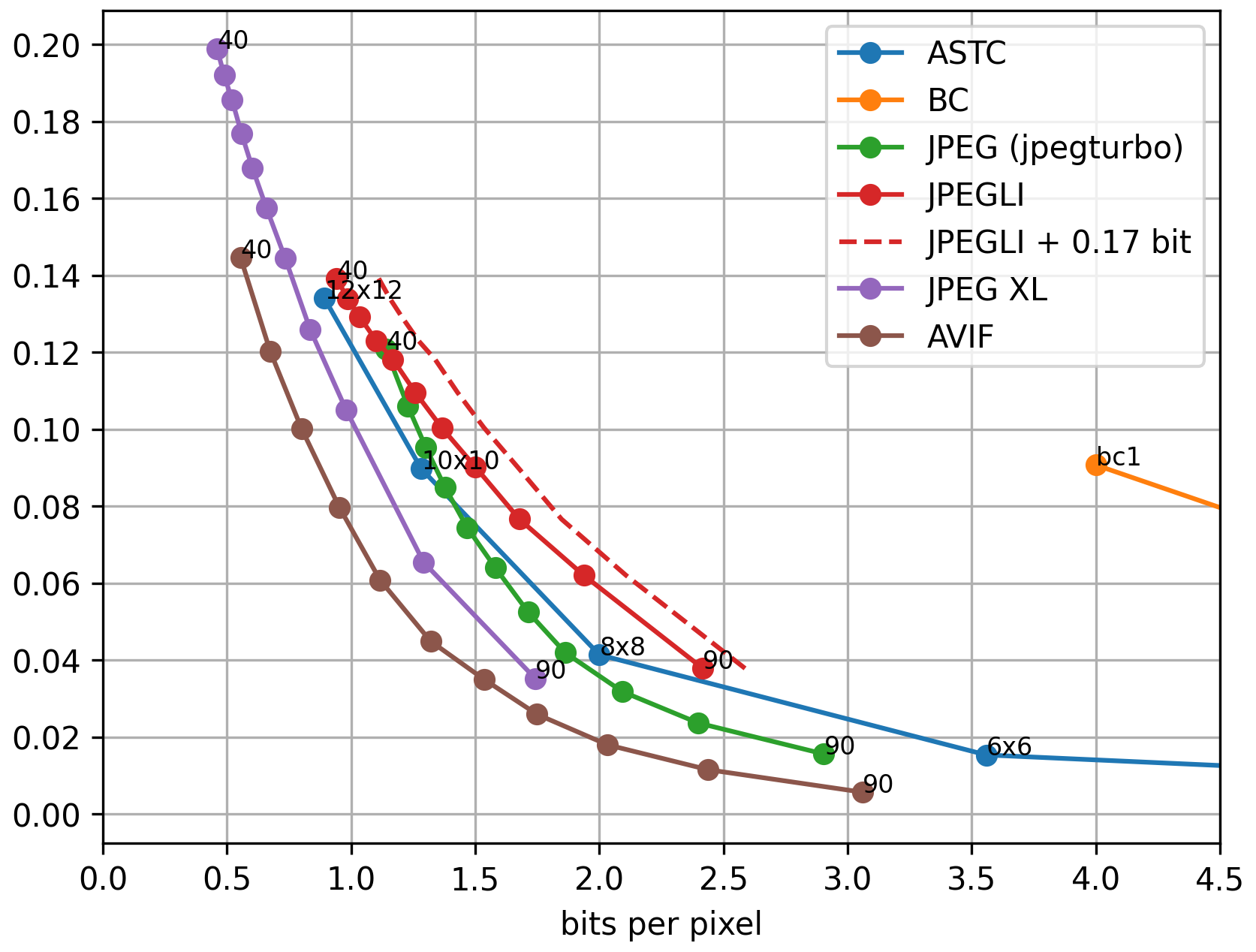}{images/qualitative/denmin_fabric_02_diff_4k.png/denmin_fabric_02_diff_4k.jpg}{34px}{0.3pt}
    \plotrow{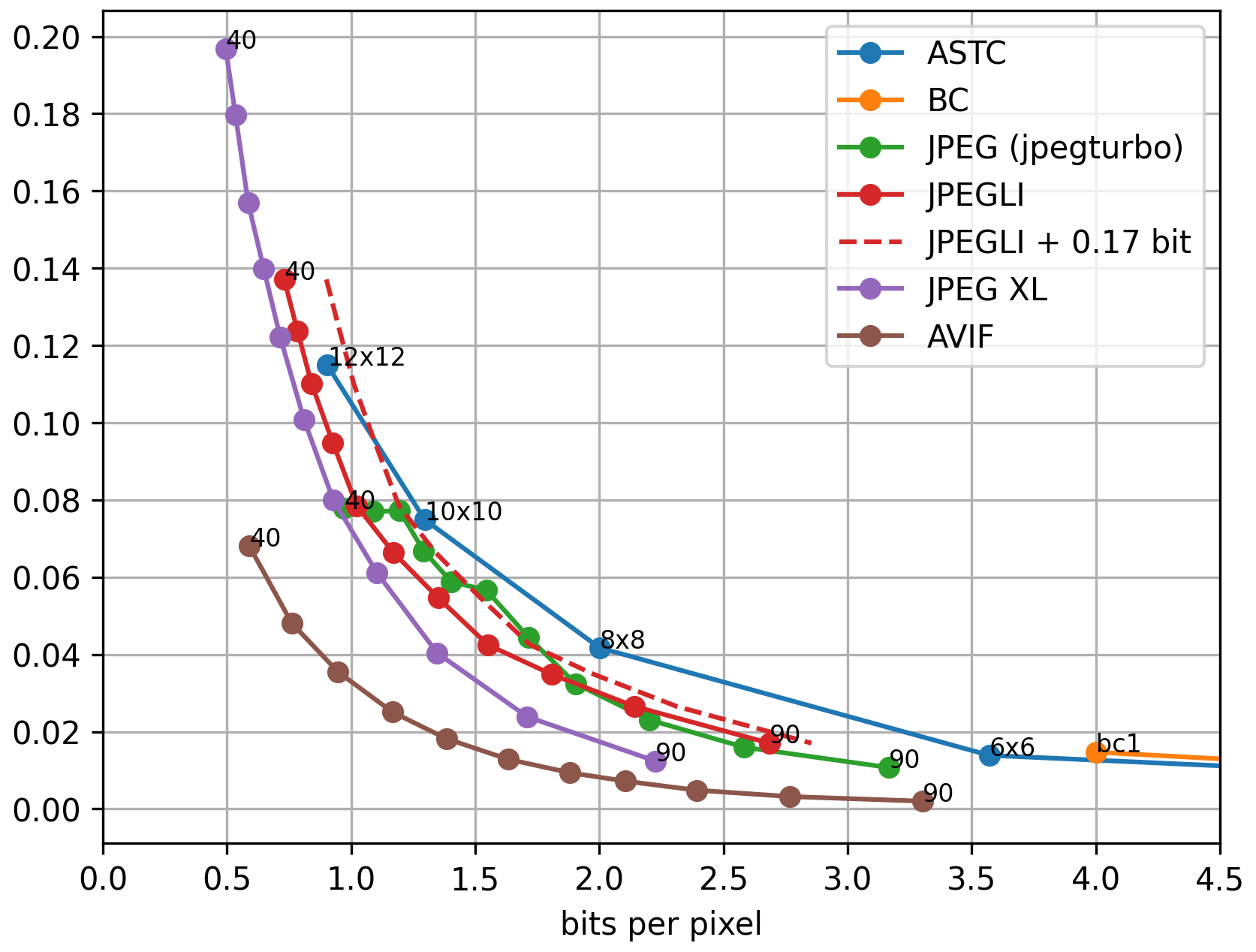}{images/qualitative/Chip005_1K-PNG_Color.png/Chip005_1K-PNG_Color.jpg}{0px}{10.3pt}
\end{tabular*}
\label{tab:appendix_quality_plots}
\end{table}

\end{document}